\documentclass[acmtog]{acmart}

\usepackage{kotex}
\usepackage{siunitx}
\usepackage{subfigure}

\usepackage{wrapfig}

\definecolor{rv}{rgb}{0,0,0}

\AtBeginDocument{%
  }

\setcopyright{acmlicensed}
\copyrightyear{2018}
\acmYear{2018}
\acmDOI{XXXXXXX.XXXXXXX}

\acmISBN{978-1-4503-XXXX-X/18/06}

\acmSubmissionID{papers\_334}

\begin{document}

\title{PhysicsFC: Learning User-Controlled Skills for a Physics-Based Football Player Controller}

\author{Minsu Kim}
\orcid{0009-0005-3798-5290}
\email{igotaspot426@gmail.com}
\affiliation{%
  \institution{Hanyang University}
  \city{Seoul}
  \country{South Korea}
}

\author{Eunho Jung}
\orcid{0009-0005-6652-5189}
\email{jho6394@hanyang.ac.kr}
\affiliation{%
  \institution{Hanyang University}
  \city{Seoul}
  \country{South Korea}
}

\author{Yoonsang Lee}
\authornote{Corresponding author.}
\orcid{0000-0002-0579-5987}
\email{yoonsanglee@hanyang.ac.kr}
\affiliation{%
  \institution{Hanyang University}
  \city{Seoul}
  \country{South Korea}
}

\begin{abstract}

We propose PhysicsFC, a method for controlling physically simulated football player characters to perform a variety of football skills--such as dribbling, trapping, moving, and kicking--based on user input, while seamlessly transitioning between these skills. 
Our skill-specific policies, which generate latent variables for each football skill, are trained using an existing physics-based motion embedding model that serves as a foundation for reproducing football motions.
Key features include a tailored reward design for the Dribble policy,
a two-phase reward structure combined with projectile dynamics-based initialization for the Trap policy,
and a Data-Embedded Goal-Conditioned Latent Guidance (DEGCL) method for the Move policy.
Using the trained skill policies, the proposed football player finite state machine (PhysicsFC FSM) allows users to interactively control the character.
To ensure smooth and agile transitions between skill policies, as defined in the FSM, we introduce the Skill Transition-Based Initialization (STI), which is applied during the training of each skill policy. 
We develop several interactive scenarios to showcase PhysicsFC's effectiveness, including competitive trapping and dribbling, give-and-go plays, and 11v11 football games, where multiple PhysicsFC agents produce natural and controllable physics-based football player behaviors.
Quantitative evaluations further validate the performance of individual skill policies and the transitions between them, using the presented metrics and experimental designs.

\end{abstract}



\begin{teaserfigure}
  \includegraphics[trim=70 158 80 120, clip, width=\textwidth]{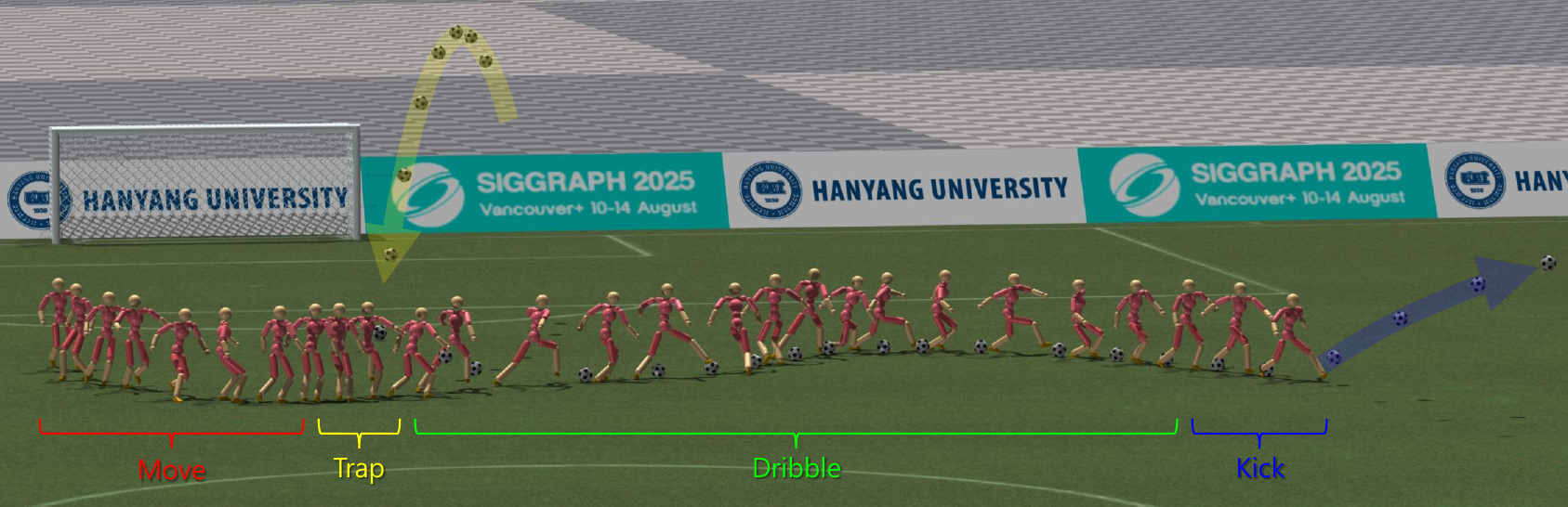}
  \caption{
  With PhysicsFC, users can control a character in a physics-simulated environment, where both the character and the ball are simulated, to perform various football skills--such as moving, trapping, dribbling, and kicking--and seamlessly transition between them.
  }
  \label{fig:teaser}
\end{teaserfigure}

\maketitle

\section{Introduction}

In football, players use a variety of skills to handle the ball with different parts of the body, excluding the arms.
Football players must control the ball while moving (dribbling), make appropriate touches to control incoming balls (trapping), and kick the ball accurately in the desired direction and with the intended force (kicking).
As all these actions involve physical interaction with the ball, it is quite challenging for an untrained person to perform these football skills well, and it becomes an even more challenging task for physically simulated characters.
Consequently, in popular football games like \textit{EA Sports FC} and \textit{eFootball}, while the ball's movement is based on physics simulation, the movements of player characters largely rely on kinematic animation.
As a result, artifacts such as foot sliding or unrealistic, abrupt movements can be observed in player character animations.
If physics-simulated characters, controlled by user input, could handle the ball as skillfully and agilely as real football players, this would significantly enhance the realism of these football games.

However, this problem is challenging for several reasons.
First, the character must not only move according to user commands but also touch the ball appropriately so it responds as the user intends.
Second, the character needs to be able to perform various skills, each with distinct purposes for handling the ball.
When dribbling, the character must keep the ball close to their feet and be able to quickly run in the desired direction.
When trapping, the character must touch incoming balls that are rolling or lobbed in from various directions and speeds, in a way that stops the ball from bouncing away.
When kicking, the character must be able to send the ball in the desired direction and with the intended force.
Additionally, the character must be capable of moving swiftly in various directions, facing different orientations, even without the ball, to quickly respond to different football situations.
Only with these skills can the character execute football plays that smoothly transition from ball possession to dribbling and then to passing or shooting.
Third, the character needs to transition between these skills smoothly and quickly.
If dribbling doesn’t immediately follow trapping, the ball would quickly be intercepted by the opponent in a real football game.
Fourth, each skill and skill transition must be user-controllable yet also capable of switching automatically based on context.

In this paper, we propose PhysicsFC, a method for controlling physically simulated football player characters to perform a variety of football skills--such as dribbling, trapping, moving, and kicking--based on user input, while seamlessly transitioning between these skills.
Our skill training follows a hierarchical framework, where each skill is managed by a dedicated skill policy trained to output latent variables for the corresponding football skill based on a user-specified goal.
These skill policies are trained using an existing physics-based motion embedding model, which provides a low-level policy capable of reproducing football motion capture data.
The latent output of each skill policy is fed into the low-level policy, enabling the character to perform various ball-handling football movements in a physics simulation, guided by the user's intent.

The key design decisions for training these skill policies play a crucial role in addressing critical aspects of their respective tasks.
Specifically, the Dribble policy is trained with a reward function that considers the ball's velocity, the distance to the ball, and the character's velocity toward the ball, enabling the character to keep the ball close to its feet while moving at the target velocity.
The Trap policy employs a two-phase reward structure (pre- and post-collision) combined with projectile dynamics-based ball-state initialization, ensuring that during training, the ball is directed to a position where the character can catch it, enabling the policy to learn how to stably receive the ball without losing control after it bounces.
The Move policy is trained using the Data-Embedded Goal-Conditioned Latent Guidance (DEGCL) method, leveraging goal-action relationships from the training data. This allows the character to mimic football player motions, maintaining diverse frontal orientations while moving in different directions and speeds.

Using the trained skill policies, the proposed football player finite state machine (PhysicsFC FSM) allows users to interactively control the character, by setting goals for each football skill and enabling transitions between them according to predefined conditions.
These transition conditions rely on user input (e.g., dribble $\rightarrow$ kick) and/or situational factors (e.g., move $\rightarrow$ dribble), depending on the skills being transitioned.
To enable smooth and agile transitions between skill policies, we introduce the Skill Transition-Based Initialization (STI), which is applied during the training of skill policies.
This method involves sampling episode initial states from simulations performed by the previous skill policy.

We develop several interactive scenarios to showcase PhysicsFC’s effectiveness.
These include competitive trapping and dribbling, give-and-go plays, and 11v11 football games, where multiple PhysicsFC agents are simulated together, showcasing the system’s ability to generate natural football player behaviors that are physics-based and user-controlled.
Additionally, we perform quantitative evaluations to assess the performance of individual skill policies and the transitions between them, using the presented metrics and experimental designs.

Our main contributions are as follows:
\begin{itemize}
    \item \textcolor{rv}{Skill policy training with tailored reward structures and initialization strategies, exemplified  by the dribble reward encouraging close ball control during high-speed movement, and the two-phase reward and the projectile-based ball initialization for robust trapping of lob and ground passes.}

    \item Data-Embedded Goal-Conditioned Latent Guidance (DEGCL), which leverages goal-action relationships in the training data to learn goal-aligned actions, enabling the Move policy to produce motions consistent with the training data across varying speeds and orientations.
    
    \item \textcolor{rv}{Skill Transition-Based Initialization (STI), which initializes episodes using intermediate states generated by preceding skill policies to support smooth and responsive transitions in real-time, user-controlled football gameplay.}
    
    \item 
    FSM-based football player controller (PhysicsFC FSM) that switches skill policies based on user input and/or surrounding conditions, enabling user-interactive control in various football scenarios.

    \item \textcolor{rv}{Comprehensive evaluation and real-time demonstrations, including competitive trapping and dribbling, give-and-go plays, and 11v11 football games, supported by quantitative metrics and experimental designs for assessing football skill policies and their transitions.}
    
\end{itemize}

\section{Related Work}

\paragraph{Physics-Based Character Control}

\textcolor{rv}{Early work on physics-based character control demonstrated that dynamic simulation combined with control algorithms could produce realistic athletic behaviors such as running, bicycling, and vaulting~\cite{hodgins_animating_1995}. 
Subsequent efforts} developed locomotion control methods based on manually crafted error feedback \cite{yin2007simbicon, coros2010generalized, lee2010data} and optimizing controller parameters \cite{wang_optimizing_2009, liu_terrain_2012}.
The introduction of deep reinforcement learning (DRL) marked a significant breakthrough in this field.
Following DeepMimic~\cite{peng2018deepmimic}, which achieved remarkable motion quality for a wide range of reference motion clips, researchers have sought to go beyond simple imitation to expand motion repertoires in diverse scenarios and enable characters to perform various tasks.
Efforts include training policies to track motions synthesized from large motion datasets, enabling the runtime reproduction of diverse motions \cite{bergamin_drecon_2019, Park2019},
and exploring the motion space around  given motion clips to extend the range of expressible actions  \cite{Lee:2021:Parameterized, chimeras22, allsteps20}.
Other approaches involve leveraging simplified physics models to expand the range of physically plausible motions while enhancing the generalization capability of learned policies \cite{kwon2020fast, kwon_adaptive_2023},
and part-wise control methods that decouple full-body control into independent body parts \cite{bae_pmp_2023,xu_composite_2023}.

Recently, physics-based motion embedding models, which learn latent representations to reproduce motions from datasets, have become a key area of focus.
Merel et al.~\shortcite{DBLP:conf/iclr/MerelHGAPWTH19} proposed a method for distilling multiple skill networks—each trained to perform a specific motion—into a single policy that acts as a low-level controller for downstream tasks.
Using this approach, they incorporated visual signals to enable full-body tasks like grasping and carrying objects \cite{merel_catch_2020}.
Peng et al.~\shortcite{ASE} introduced Adversarial Skill Embedding (ASE), which uses GAIL to learn a reusable skill prior that can replicate diverse motions.
This prior allow high-level controllers to be effectively trained to handle complex tasks.
Building on ASE, C-ASE~\cite{dou_case_2023} incorporated conditional control to enable more precise motion generation tailored to specific tasks.
CALM~\cite{CALM}, on the other hand, eliminated the need for mutual information maximization used in ASE and instead focused on training conditional adversarial latent models to enhance the diversity and goal-oriented nature of motion generation.
Won et al.~\shortcite{won_physics-based_2022} introduced a conditional VAE-based approach that trains a motor decoder in combination with a world model, providing low-level control policies capable of supporting a wide range of downstream tasks.
Yao et al.~\shortcite{yao_controlvae_2022} explored VAE-based motion generation by jointly learning state-conditional priors and world models, enabling the effective training of high-level control policies that integrate multiple skills.
Zhu et al.~\shortcite{zhu_neural_2023} adopted a VQ-VAE approach to learn a discrete latent space that mitigates mode collapse and improves the stability and expressiveness of latent representations.
MoConVQ~\cite{yao_moconvq_2024} introduced a framework that uses VQ-VAE and world models to learn latent representations from large motion datasets, integrating multiple physics-based control tasks such as text-to-motion and interaction control into a unified system.
These studies highlight the importance of structured latent space learning in developing high-level skill policies for physics-based systems.
Building on these advancements, we aim to develop a realistic, user-controllable physics-based football player controller by training football skill policies and integrating them into a unified framework, utilizing CALM~\cite{CALM} as a latent model.

\paragraph{Physics-Based Sports Motion Control}

Physics-based methods for sports motion control focus on synthesizing dynamic and realistic movement strategies for virtual characters.
Si et al.~\shortcite{si_realistic_2014} proposed a biomechanical swimming model with biomimetically inspired motor control system based on Central Pattern Generators (CPGs) to generate natural swimming motions.
Liu et al.~\shortcite{liu_learning_2018} proposed a physics-based basketball dribbling controller that uses trajectory optimization and DRL for ball-handling.
Yin et al.~\shortcite{jump2021} explored diverse jumping strategies through DRL, allowing characters to discover natural and varied athletic jumping behaviors.
Won et al.~\shortcite{won_control_2021} proposed a two-step approach to learning control policies for two-player sports like boxing and fencing, focusing on basic skills and opponent-based strategies.
Zhang et al.~\shortcite{zhang_learning_2023} focused on learning physically plausible tennis skills from broadcast video data, enabling realistic shotmaking and rallies.
Wang et al.~\shortcite{wang_strategy_2024} proposed a hierarchical system for physics-based table tennis, combining skill and strategy learning to enable effective decision-making in dynamic environments.

Specifically, football has been a focus of physics-based simulation methods applied to skill execution, team coordination, and humanoid robot gameplay.
DeepLoco~\cite{deeploco} demonstrated a dribbling task where a character moves a ball to a target location using its feet.
However, the motion was more akin to walking with occasional foot taps, lacking the agility of a football player’s dribbling.
Hong et al.~\shortcite{kaist_soccer} demonstrated more realistic football dribbling and shooting by combining data-driven motion prediction with Model Predictive Control (MPC).
While their approach produced natural dribbling motions in a physics simulation environment, it struggled to maintain close ball control, with the character kicking the ball far and chasing it.
\textcolor{rv}{Peng et al.~\cite{AMP} demonstrated a dribbling task using adversarial motion priors, where the character moves a ball toward a target. However, the focus was on reproducing stylistic behaviors from unstructured motion data rather than achieving fine-grained ball control.}
Liu et al.~\shortcite{deepmind_soccer} presented a framework for transitioning from motor control to team play, where simulated humanoid agents coordinate their behaviors and learn to perform complex team-based 2v2 football scenarios.
While their approach achieves team-level coordination, the character movements, such as dribbling, remain less smooth and natural compared to those of real human players.
Additionally, Xie et al.~\shortcite{soccer22} proposed a system that trains physics-based controllers for diverse soccer juggling skills and smooth transitions between them using a layer-wise mixture-of-experts architecture, 
while Haarnoja et al.\shortcite{haarnoja_learning_2024}  trained humanoid robots for 1v1 soccer using a two-stage pipeline, combining skill distillation and multi-agent self-play to enable dynamic skills and opponent anticipation, with safe zero-shot transfer to real robots.

Our approach addresses several limitations of prior studies by controlling physics-based characters to perform football motions that are both natural and closely resemble the movements of real players.
Our controller replicates an elite player’s dribbling by keeping the ball close to the character’s feet and introduces a novel trapping skill to stop and control a moving ball, which has been underexplored in previous research.
While prior studies often focus on individual football skills, we propose a learning framework that not only trains multiple user-controllable skills but also facilitates smooth transitions between these skills, with metrics and experimental designs devised for quantitative evaluation.
Finally, our framework integrates these capabilities into a unified control system, providing a physics-based football gameplay experience that captures the fluidity of real-world football.

\section{Overview}
\label{sec:overview}

\begin{figure}
  \centering
  \includegraphics[trim=65 0 65 0, clip, width=1\linewidth]{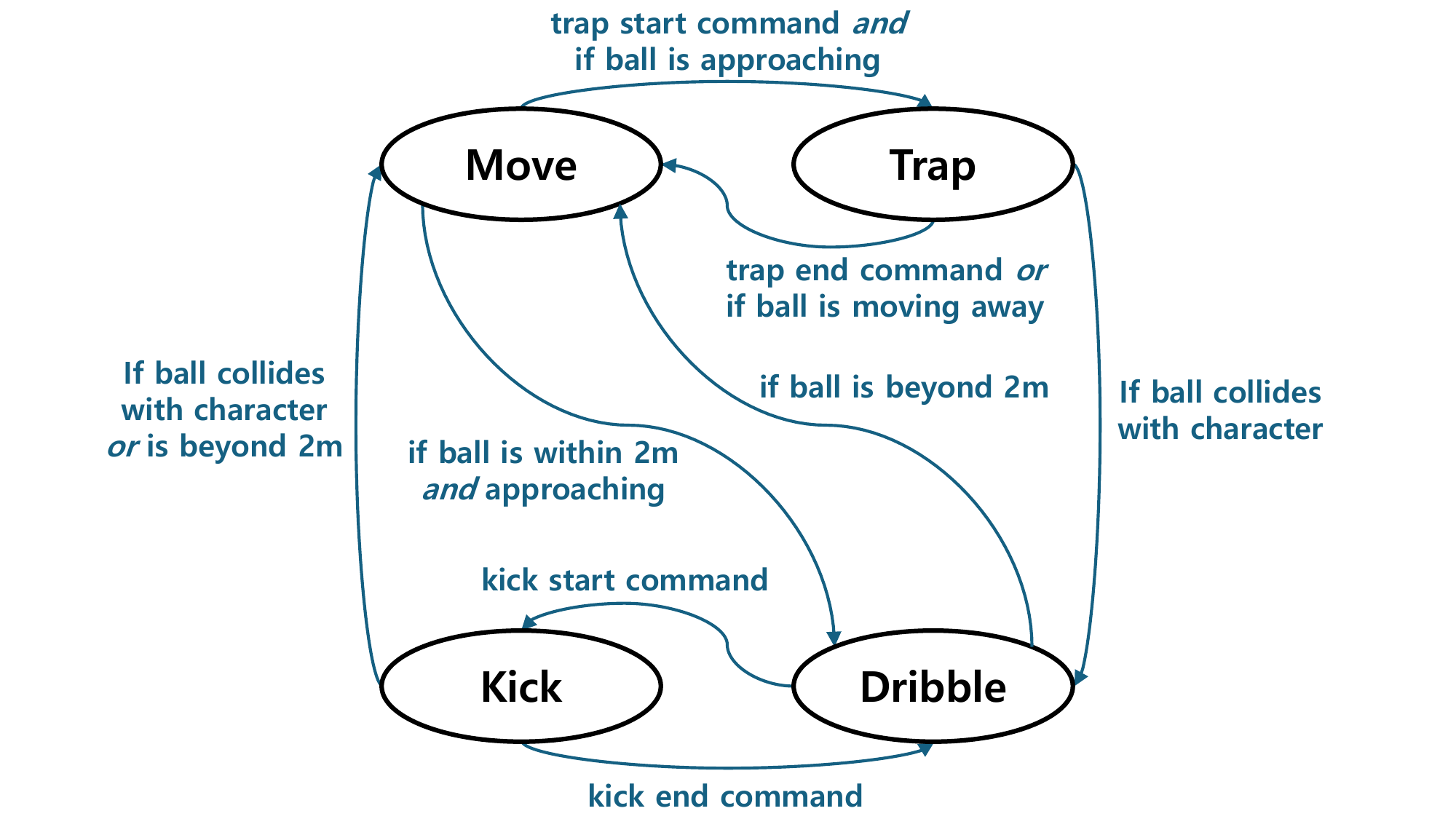}
  \caption{PhysicsFC FSM.}
\label{fig:fsm}
\end{figure}

In our method, each football skill is executed by a separate policy.
The Dribble, Trap, Move, and Kick policies are designed with distinct goal inputs, rewards, episode initialization, and termination routines.
These skill policies are trained to operate with a shared low-level policy from a physics-based motion embedding model, which is independently trained using football motion capture data.
The output of each skill policy is a latent vector $\mathbf z$, which serves as input to the low-level policy.
Note that the ball is simulated only during high-level skill policy training, not during low-level policy training.
Detailed descriptions of each skill policy are provided in Section~\ref{sec:skills}.

The physics-based motion embedding model learns a low-level policy capable of physically reproducing motions from the motion dataset, as well as the latent space where the input $\mathbf z$ to the low-level policy is defined.
When the low-level policy receives a particular $\mathbf z$ along with the current character state and outputs a low-level action to the physics simulation, the simulated character performs a motion corresponding to $\mathbf z$, consistent with the motions in the dataset.
We use CALM~\cite{CALM} as the physics-based motion embedding model, and further details are provided in Appendix~\ref{sec:appd-calm}.

For runtime control, we propose PhysicsFC FSM, a football player FSM that integrates the learned football skill policies and the predefined transition conditions between them (Figure~\ref{fig:fsm}).
At each moment, the user's commands serve either as goal inputs for the skill policy corresponding to the current FSM state or as triggers to initiate a state transition.
This enables the user to control the character to perform physics-based football gameplay, progressing from ball possession to dribbling, and then to passing or shooting.
Detailed explanations are provided in Section~\ref{sec:fsm}.

\section{Football Skill Policies}
\label{sec:skills}

Each skill policy takes as input the character state and ball state, along with the goal for each skill described in the subsections below, and outputs a latent $\mathbf z$ to be fed into the low-level policy. Detailed descriptions of the character state and ball state can be found in Appendix~\ref{sec:appd-character-ball-states}.

\paragraph{\textbf{Skill Transition-Based State Initialization (STI)}}

\begin{figure}
  \centering
  \includegraphics[trim=215 30 215 30, clip, width=.9\linewidth]{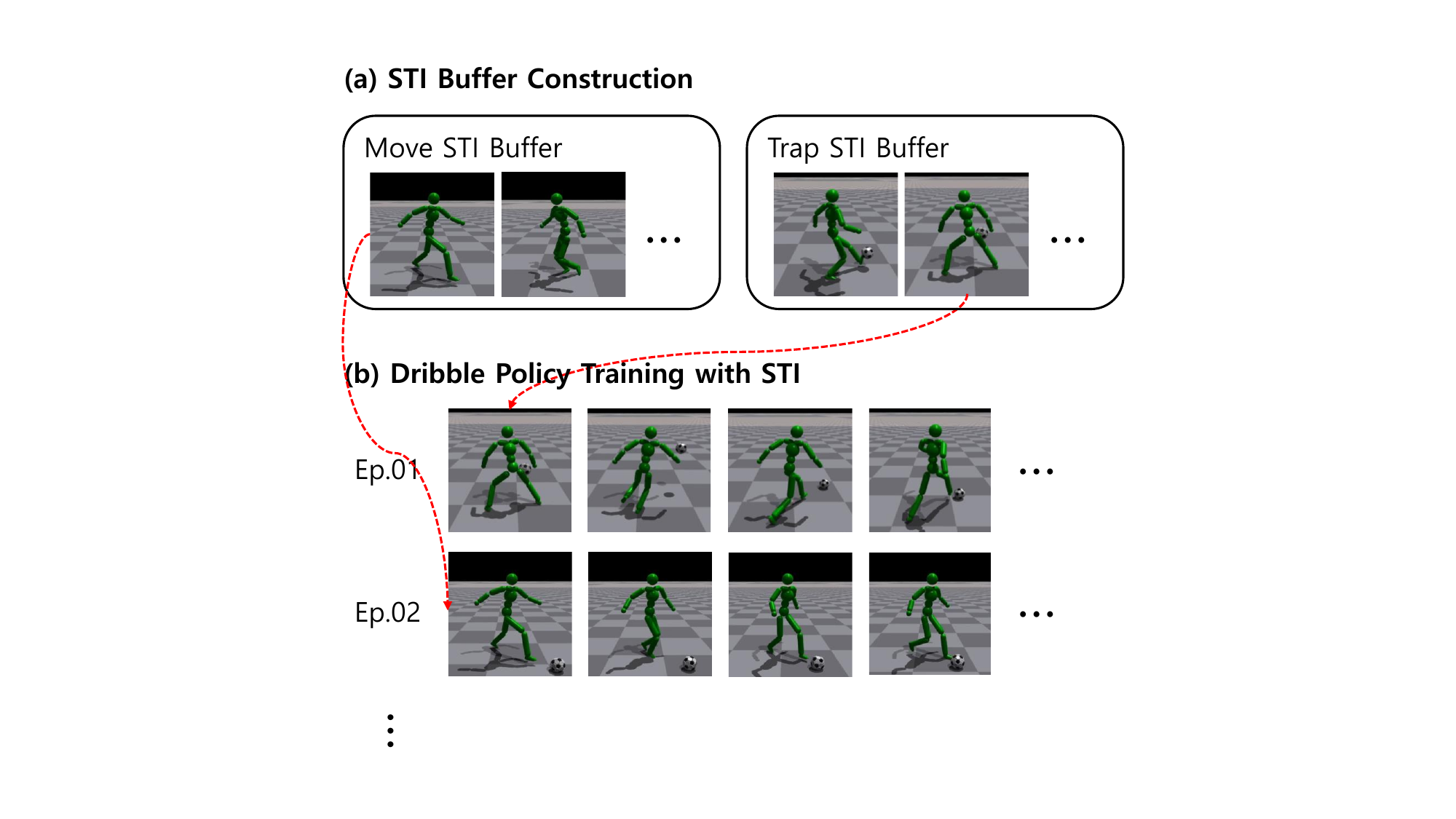}
  \caption{
  Example of Dribble Policy Training with Skill Transition-Based State Initialization (STI):
  (a) Numerous episodes are simulated using trained skill policies, and the character and ball states are stored in STI buffers for each skill.
  (b) During Dribble policy training, half of the episodes are initialized with states randomly sampled from the Trap STI buffer, while the other half are initialized from the Move STI buffer.
  Through these episodes, the Dribble policy learns to initiate dribbling quickly in various situations, both while moving and immediately after trapping.
  }
\label{fig:sti}
\end{figure}

For a football player to effectively bypass opponents and create opportunities during a match, it is crucial to transition quickly between football skills while maintaining smooth control of the ball.
For example, the player may need to start dribbling immediately after trapping the ball or perform a sudden kick during a dribble.

To achieve this, each skill policy is trained using a technique called Skill Transition-Based State Initialization (STI).
STI initializes episodes by leveraging skill policies that represent the previous states in our PhysicsFC FSM (Figure 2).
Specifically, simulated data is generated for each trained skill policy, with randomly selected goal inputs within the same range as during policy training.
These data are stored in skill-specific STI buffers, capturing character and ball states at moments where FSM-defined transitions can occur.
For instance, for the Dribble policy, states are saved at every reinforcement learning (RL) step since a kick start command can be issued at any time during dribbling.
For the Trap skill, states are saved only when the ball collides with the character, as this is the moment when the transition to Dribble occurs.
STI is not applied to the Move policy because transitions to Move do not involve continuous ball handling after the transition.
During training, the initial state of each episode is randomly sampled from the STI buffer of the skill corresponding to the previous FSM state.
This ensures that each skill policy learns to transition seamlessly from its predecessor policy at runtime.

\textcolor{rv}{Unlike previous approaches that primarily leverage terminal states of preceding policies~\cite{konidaris_skill_2009,chen2023sequential}, our method captures intermediate transition points throughout skill execution.
This design choice better aligns with the requirements of interactive football gameplay, where transitions must occur fluidly based on dynamic and unpredictable user inputs.
}

\paragraph{Foot Collision Mesh}

\begin{wrapfigure}{r}{0.17\columnwidth}
      \includegraphics[trim=0 0 0 0, clip, width=0.2\columnwidth]{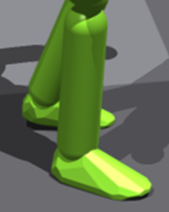}
  \caption{
    Foot collision mesh.
    }
  \label{fig:foot}
\end{wrapfigure}

We found that the shape of the foot collision mesh has a noticeable effect on dribbling and kicking performance.
When using the commonly employed box-shaped foot, the Dribble policy failed to learn how to agilely adjust the dribbling direction toward the desired target, and the Kick policy produced kicking motions where the velocity of the kicked ball showed a noticeable deviation from the target velocity.
To improve the performance, we applied a football boots mesh file as the foot collision mesh, which is automatically converted into a convex hull by the physics engine and used as the collision shape (Figures~\ref{fig:foot}).
The character with a football boots-shaped foot successfully learned to perform dribbling and kicking, demonstrating the importance of using an appropriately shaped collision mesh.

\subsection{Dribble}
\label{sec:dribble}
Dribbling is a skill where a player moves while controlling the ball with their feet, commonly used to advance and create scoring opportunities or to set up passing lanes for teammates.
Keeping the ball close to the feet is essential, as it allows the player to react quickly to sudden changes, maintain precise control, and hold possession securely without easily losing the ball to opponents.
Our Dribble policy is trained to enable the character to keep the ball close to their feet while moving it at the user-specified target velocity.

\paragraph{Goal Input}
The Dribble policy takes the target dribble velocity, $\hat{\mathbf v}^{\mathrm{drib}}_t \in \mathbb{R}^2$, as its goal.
During training, $\hat{\mathbf v}^{\mathrm{drib}}_t$ is sampled separately for speed and direction: the target speed is uniformly sampled from $[0, \SI{7}{m/s}]$, while the target direction is uniformly sampled from $[0, 360^\circ]$.
The goal is set randomly at the start of each episode and reassigned at a random time between $[5, \SI{6.5}{s}]$, with the maximum episode length being 10 seconds. The actual policy input is the target velocity $\hat{\mathbf v}^{\mathrm{drib}}_{t\{c\}}$, expressed in the character coordinate system, described in Appendix~\ref{sec:appd-character-ball-states}.

\paragraph{Reward}

The reward for the Dribble policy is defined as follows (Figure~\ref{fig:dribble-reward}):

\begin{figure}
  \centering
  \includegraphics[trim=100 0 50 0, clip, width=1\linewidth]{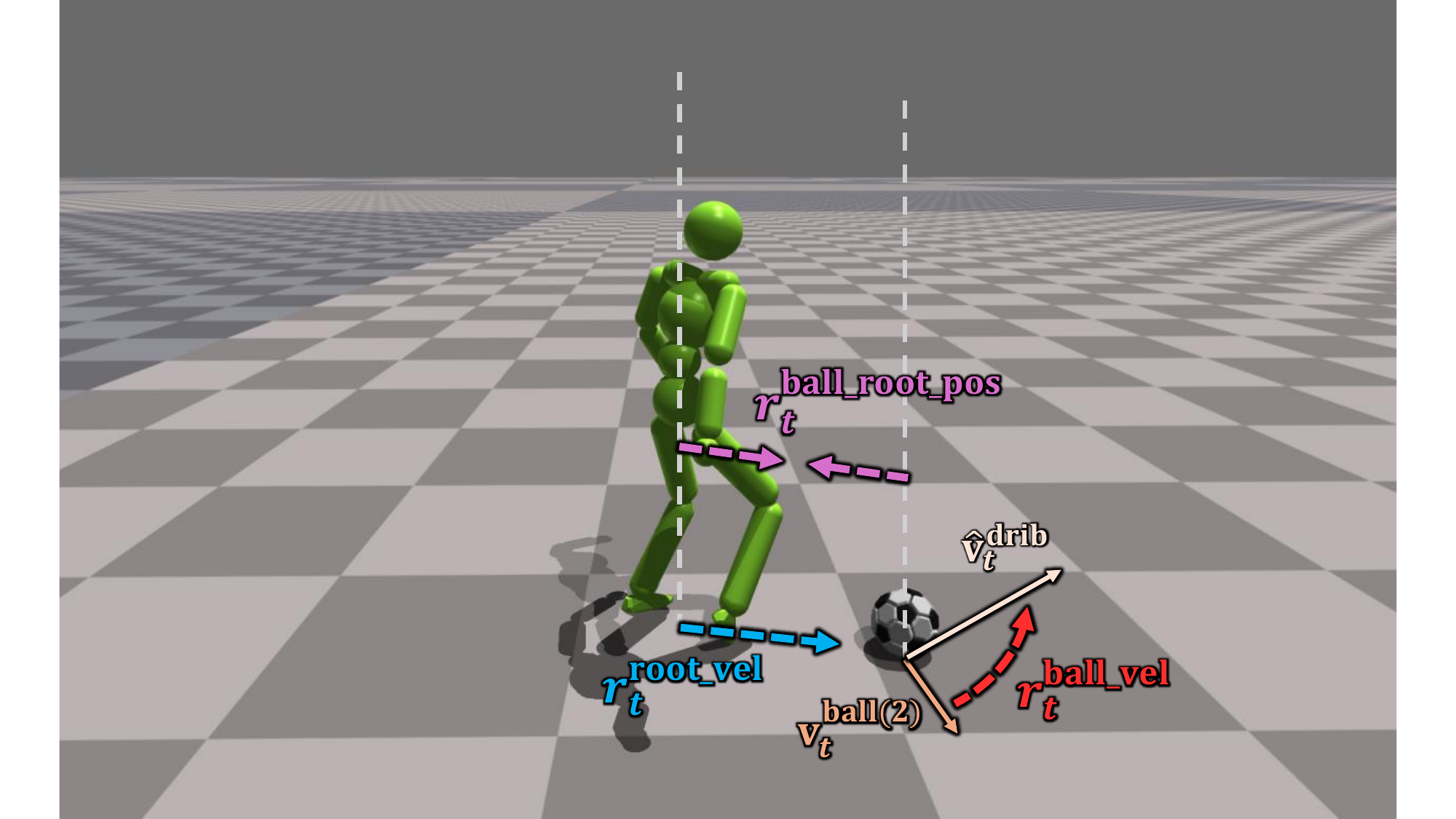}
  \caption{Visualization of Dribble reward.}
\label{fig:dribble-reward}
\end{figure}

\begin{equation}
\begin{aligned}
r^{\mathrm{drib}}_t = \ 
& 0.6 \ r^{\mathrm{ball\_vel}}_t \left( \hat{\mathbf v}^{\mathrm{drib}}_t, \mathbf v^{\mathrm{ball(2)}}_t \right) + \\
& 0.2 \ r^{\mathrm{ball\_root\_pos}}_t \left( \mathbf x^{\mathrm{root(2)}}_t, \mathbf x^{\mathrm{ball(2)}}_t \right) + \\
& 0.2 \ r^{\mathrm{root\_vel}}_t \left( \hat{\mathbf v}^{\mathrm{drib}}_t, \mathbf v^{\mathrm{root(2)}}_t, \mathbf x^{\mathrm{root(2)}}_t, \mathbf x^{\mathrm{ball(2)}}_t \right).
\label{eq:r_drib}
\end{aligned}
\end{equation}

The term $r^{\mathrm{ball\_vel}}_t(\cdot)$ encourages the horizontal velocity of the dribbled ball 
$\mathbf v^{\mathrm{ball(2)}}_t$ to match the target dribble velocity.
This term enables the Dribble policy to learn to dribble the ball in the target direction and at the target speed.
The term $r^{\mathrm{ball\_root\_pos}}_t(\cdot)$ is designed to encourage the character to keep the ball close to its foot while dribbling, where $\mathbf x^{\mathrm{root(2)}}_t$ and $\mathbf x^{\mathrm{ball(2)}}_t$ represent the horizontal positions of the character's root (pelvis) and the ball, respectively.
Since the character's feet move back and forth repeatedly during dribbling, using the actual foot positions for reward calculation makes learning the policy challenging.
As a simple and effective alternative, we designed the reward to minimize the horizontal distance between the character's root and the ball, enabling the policy to perform dribbling that keeps the ball close to the foot.
The term $r^{\mathrm{root\_vel}}_t(\cdot)$ guides the character's root (pelvis) to move toward the ball's current horizontal position at the target dribble speed,
where $\mathbf v^{\mathrm{root(2)}}_t$ represents the horizontal velocity of the character's root.
This helps the policy learn to dribble while maintaining control of the ball.
Note that both $r^{\mathrm{ball\_vel}}_t(\cdot)$ and $r^{\mathrm{root\_vel}}_t(\cdot)$ are normalized by the target speed (NTS)
to ensure consistent reward signals regardless of the target speed's magnitude, facilitating stable learning across different speed conditions.
All positions and velocities used in the calculation of Equation (1) are expressed in the global coordinate system. For detailed descriptions of each term and the NTS, please refer to Appendix~\ref{sec:appd-dribble-reward}.

\paragraph{Episode Initialization}
In our PhysicsFC FSM, the Dribble policy transitions from the Trap, Move, and Kick policies (Figure~\ref{fig:fsm}).
Among these, transitions from the Kick policy occur when the user cancels a kick, leading the character to be in a posture similar to that of the Move skill.
For this reason, during the training of the Dribble policy, episodes are initialized using only the trained Trap and Move policies through STI.
Specifically, episodes are initialized with state samples randomly selected from the Trap STI buffer 50\% of the time and from the Move STI buffer the other 50\%. 
When initialized with a Trap sample, both the character and the ball states are set to the values from the sampled state. When initialized with a Move sample, only the character state is taken from the sample, while the ball’s position is randomly placed on the ground within a radius of \SI{1}{m} around the character’s root, and its velocity is initialized in a random direction parallel to the ground, with a speed between \SI{0}{m/s} and \SI{1}{m/s}.

\paragraph{Episode Termination}
If the horizontal distance between the ball and the character’s root exceeds \SI{3}{m} during an episode, the episode is terminated early.
It was observed that the use of this early termination significantly impacts the successful training of the Dribble policy.
If the episode is not terminated early, it ends after 10 seconds.

\subsection{Trap}
\label{sec:trap}
Trapping is a skill used to control a ball, whether rolling on the ground (ground pass) or falling from the air (lob pass), using body parts such as the feet, legs, or chest.
This technique ensures stable ball control and prepares the player for subsequent actions like passing, dribbling, or shooting.
Our Trap policy is trained to reliably handle both ground passes and lob passes by making contact with the specified body part.

\paragraph{Goal Input}

The trap policy takes a one-hot vector input representing the body part used to touch the ball.
During training, for a lob pass, one of six body parts (head, torso, either lower leg, or either foot) is randomly selected as input, while for a ground pass, one of the two feet is randomly selected.

\paragraph{Reward}
The reward for the Trap policy is defined as follows (Figure~\ref{fig:trap-reward}):

\begin{figure}
  \centering
  \subfigure[Pre-collision phase]{
  \includegraphics[trim=330 70 210 50, clip, width=.475\linewidth]{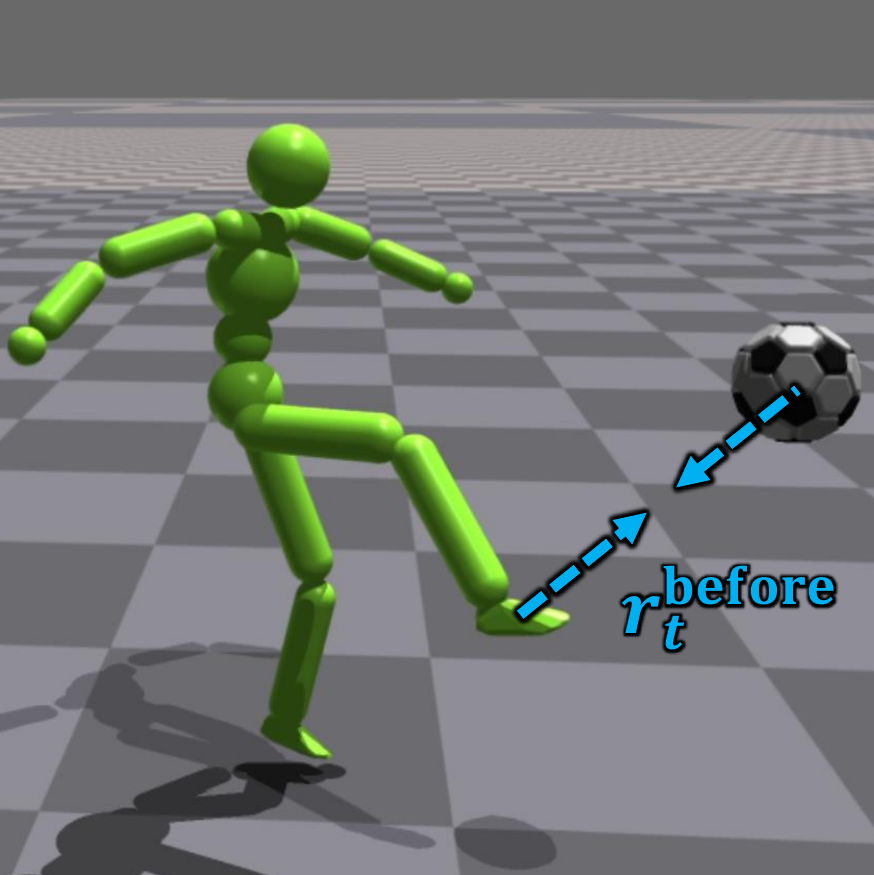}
  }
  \subfigure[Post-collision phase]{
  \includegraphics[trim=320 70 220 50, clip, width=.475\linewidth]{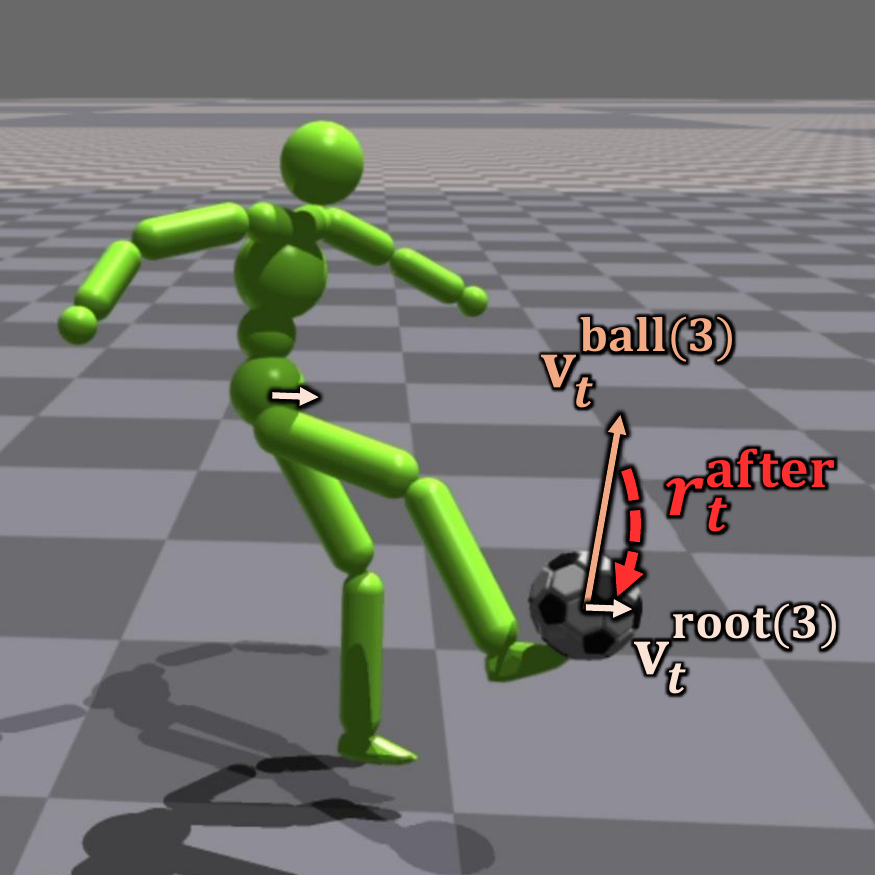}
  }
  \caption{Visualization of Trap reward.}
\label{fig:trap-reward}
\end{figure}

\begin{equation}
\begin{aligned}
r^{\mathrm{trap}}_t = 
\begin{cases}
r^{\mathrm{before}}_t = \mathrm{exp} \left(-10 \left\| \mathbf x^{\mathrm{ball(3)}}_t - \mathbf x^{\mathrm{body}}_t \right\| ^2 \right), &\text{if } t \leq t_c \\
r^{\mathrm{after}}_t = \mathrm {exp} \left(-10 \left\| \mathbf v^{\mathrm{ball(3)}}_t - \mathbf v^{\mathrm{root(3)}}_t \right\| ^2 \right), &\text{otherwise}
\end{cases}
\label{eq:r_trap}
\end{aligned}
\end{equation}
where $t_c$ represents the moment when the ball first collides with the character,
$\mathbf x^{\mathrm{ball(3)}}_t$ denotes the 3D position of the ball,
$\mathbf x^{\mathrm{body}}_t$ refers to the position of the body part specified by the one-hot vector input as the part used to touch the ball,
and $\mathbf v^{\mathrm{ball(3)}}_t$ and $\mathbf v^{\mathrm{root(3)}}_t$ indicate the 3D velocities of the ball and the character root, respectively, all expressed in global coordinates.
Equation~\ref{eq:r_trap} is divided into two phases: before and after the collision time $t_c$.
The pre-collision reward, $r^{\mathrm{before}}_t$, encourages the specified body part to make contact with the ball by assigning higher values as the distance between the body part and the ball decreases.
The post-collision reward, $r^{\mathrm{after}}_t$, ensures stable ball control by effectively absorbing the ball's momentum upon impact, assigning higher values as the ball's relative velocity to the character root approaches $\mathbf 0$.
This post-collision reward is evaluated over a very short duration (1/6 second) after the collision.

\paragraph{Episode Initialization}

\begin{figure}
  \centering
    \subfigure[Lob pass]{
       \includegraphics[trim=30 120 30 120, clip, width=1.\columnwidth]{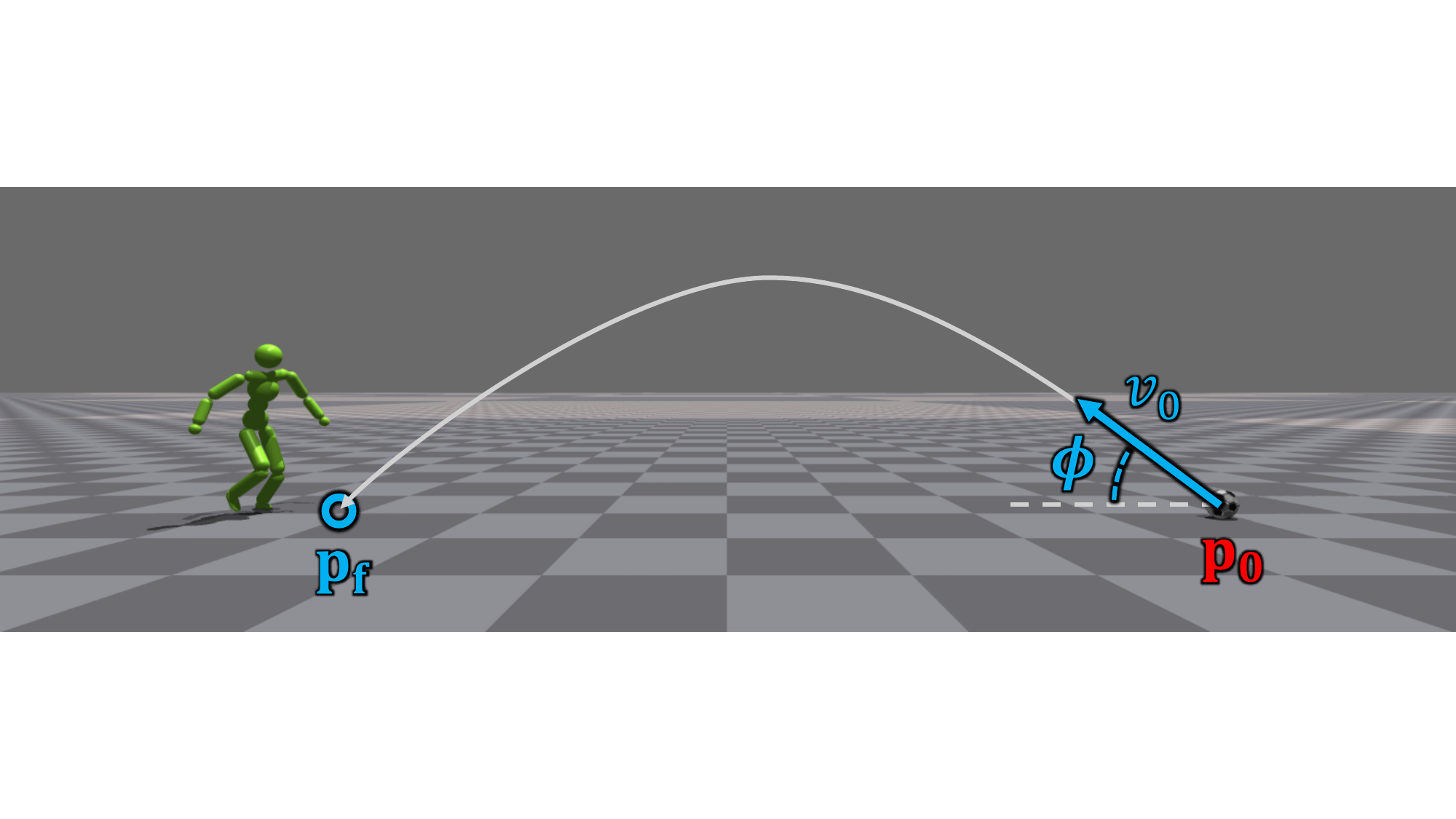}
        \label{fig:trap-init-lob}
    }
    \subfigure[Ground pass]{
       \includegraphics[trim=30 120 30 120, clip, width=1.\columnwidth]{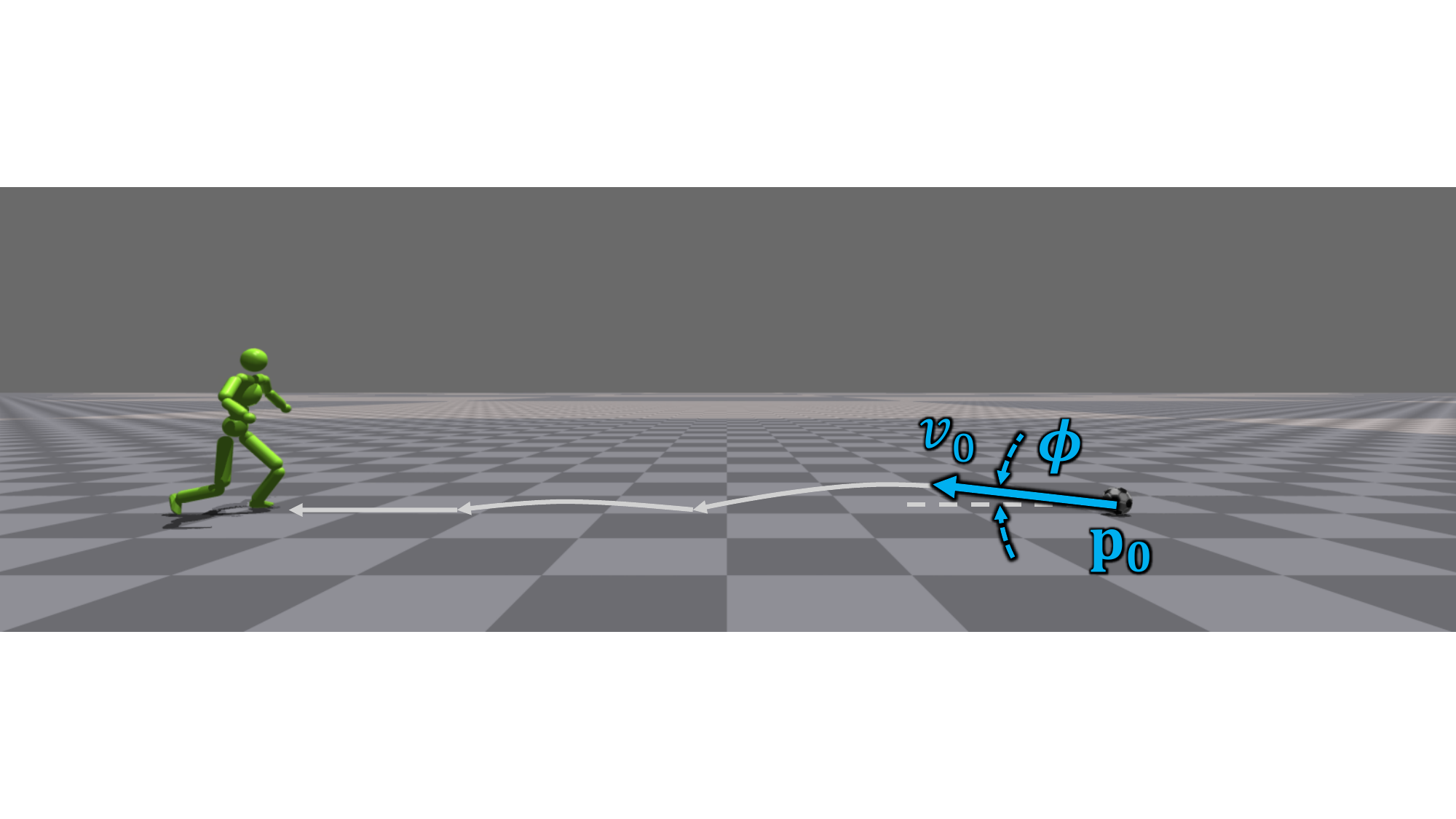}
        \label{fig:trap-init-ground}
    }    
  \caption{
  Ball state initialization for Trap policy training.
  The horizontal direction of the incoming ball is randomly determined, so this figure illustrates the motion within a vertical plane perpendicular to the ground.
  Blue symbols represent randomly assigned values, while red symbols indicate analytically calculated values based on them.
  (a) Lob pass: The ball's initial state is determined by calculating its initial position $\mathbf p_0$, ensuring it is launched with a randomly assigned initial speed $v_0$ and vertical launch angle $\phi$, and lands at a randomly specified landing position $\mathbf p_\mathrm f$.
  (b) Ground pass: The ball's initial state is determined using a randomly assigned initial speed $v_0$, initial position $\mathbf p_0$, and vertical launch angle $\phi$.
  }
 \label{fig:trap-initialization}
\end{figure}

A trap is a skill used to gain possession of the ball when the character does not currently have it.
As such, our Trap policy transitions solely from the move policy (Figure~\ref{fig:fsm}).
During Trap policy training, episodes are initialized by randomly sampling the character’s state from the Move STI buffer.
Since the Move STI buffer does not include the state of the ball, the ball is initialized differently for lob passes and ground passes, as described below.

For \textbf{lob passes}, it is essential to initialize the ball's state so that it lands within a reachable position for the character.
If the ball consistently lands too far from the character, the character cannot interact with it, preventing the policy from learning how to trap the ball.
To address this, we calculate the ball's initial position so that, when launched with a randomly selected speed and direction, it lands at a randomly chosen target location near the character (Figure~\ref{fig:trap-init-lob}).
Specifically, the ball's initial speed $v_0$ is uniformly sampled from [10, \SI{30}{m/s}], and its initial angular velocity is uniformly sampled from [0, \SI{80}{rad/s}] about a randomly chosen axis of rotation.
The landing position $\mathbf p_\text f$ is randomly sampled within a semicircular area of radius \SI{1}{m}, centered around a point offset from the character's initial position by the distance it would travel due to its initial horizontal velocity during the ball's flight time, $s$. The semicircle is oriented along the character's initial movement direction, spanning $\pm 45^\circ$ from that direction.
Both the initial and landing heights are set to the radius of the ball.
When the initial and landing positions are at the same height, we can derive the relationship between the vertical launch angle $\phi$ and the horizontal distance $d$ between them, as well as the ball's flight time $s$ (Appendix~\ref{sec:appd-ball-distance-angle}).
We uniformly sample $\phi$ from [10, $45^\circ$], calculate $d$ and $s$ using the derived relationship, and determine the ball's initial position $\mathbf p_0$ by offsetting $d$ in a random direction from $\mathbf p_\text f$.
The horizontal launch angle is determined by $\mathbf p_0$ and $\mathbf p_\text f$.

For \textbf{ground passes}, since the ball may collide with the ground multiple times before reaching the character, the ball's initial speed, initial position, and vertical launch angle are all randomly selected to initialize its state.
Specifically, the initial speed $v_0$ is uniformly sampled from the same range as for lob passes.
The initial position $\mathbf p_0$ is calculated based on a randomly chosen target point, determined in the same manner as the landing position for lob passes.
The initial position is offset in a random horizontal direction from the target point by a distance proportional to the sampled initial speed, selected within the range of [15, \SI{45}{m}].
The initial vertical height is uniformly sampled between the radius of the ball and \SI{50}{cm}.
The horizontal launch angle is set to align with the direction from the initial position to the target point, while the vertical launch angle $\phi$ is uniformly sampled from [0, $10^\circ$].

\paragraph{Episode Termination}

In football, intentional contact between the ball and the arm or hand below the shoulder is considered a handball foul.
To ensure the character learns to trap the ball without committing a handball, episodes are immediately terminated if the ball touches the character's hand, forearm, or upper arm.
Additionally, for a lob pass, if the ball touches the ground before reaching the character, or for a ground pass, if the ball passes beyond the character, the episode is immediately terminated, as this is considered a trapping failure.
In such cases, the policy does not receive any reward from $r^{\mathrm{after}}_t$, encouraging it to avoid these  situations.
For collisions involving other parts of the character's body, episodes end 1/6 seconds after the collision.
The maximum duration of an episode is 10 seconds.

\subsection{Move}
In this paper, we use the term "move" to describe all movement actions performed by the character when not in possession of the ball.
Our Move policy is trained to enable the character to walk, run, or move sideways or backward, following the user-specified target velocity and facing direction.

\paragraph{Goal Input}
The Move policy takes the target movement velocity $\hat{\mathbf v}^{\mathrm{move}}_t \in \mathbb{R}^2$ and the target facing direction $\hat{\mathbf d}^{\mathrm{face}}_t \in \mathbb{R}^2$, a unit vector, as goal inputs.
During training, for "general" episodes, the magnitude of $\hat{\mathbf v}^{\mathrm{move}}_t$ is uniformly sampled from [0, \SI{7}{m/s}], and the directions of $\hat{\mathbf v}^{\mathrm{move}}_t$ and $\hat{\mathbf d}^{\mathrm{face}}_t$ are uniformly sampled from [0, $360^\circ$].
Similar to the Dribble policy (Section~\ref{sec:dribble}), the goal is set at the start of each episode and updated at a random time between [5, \SI{6.5}{s}].
The actual inputs to the policy are $\hat{\mathbf v}^{\mathrm{move}}_{t\{c\}}$ and $\hat{\mathbf d}^{\mathrm{face}}_{t\{c\}}$, expressed in the character coordinate system.

\paragraph{\textbf{Data-Embedded Goal-Conditioned Latent Guidance (DEGCL)}}

If the move policy is trained solely using a task reward calculated based on the differences between the character's current and desired movement velocity and facing direction, the policy may fail to fully utilize diverse movement motions (e.g., sideways walk, backward walk, etc.) present in the motion data used for training.
This occurs because the task reward can be sufficiently maximized without utilizing these motions, resulting in a learned policy that achieves the user-specified goal but produces unnatural character movements.

\begin{figure}
  \centering
  \includegraphics[trim=280 10 250 0, clip, width=.95\linewidth]{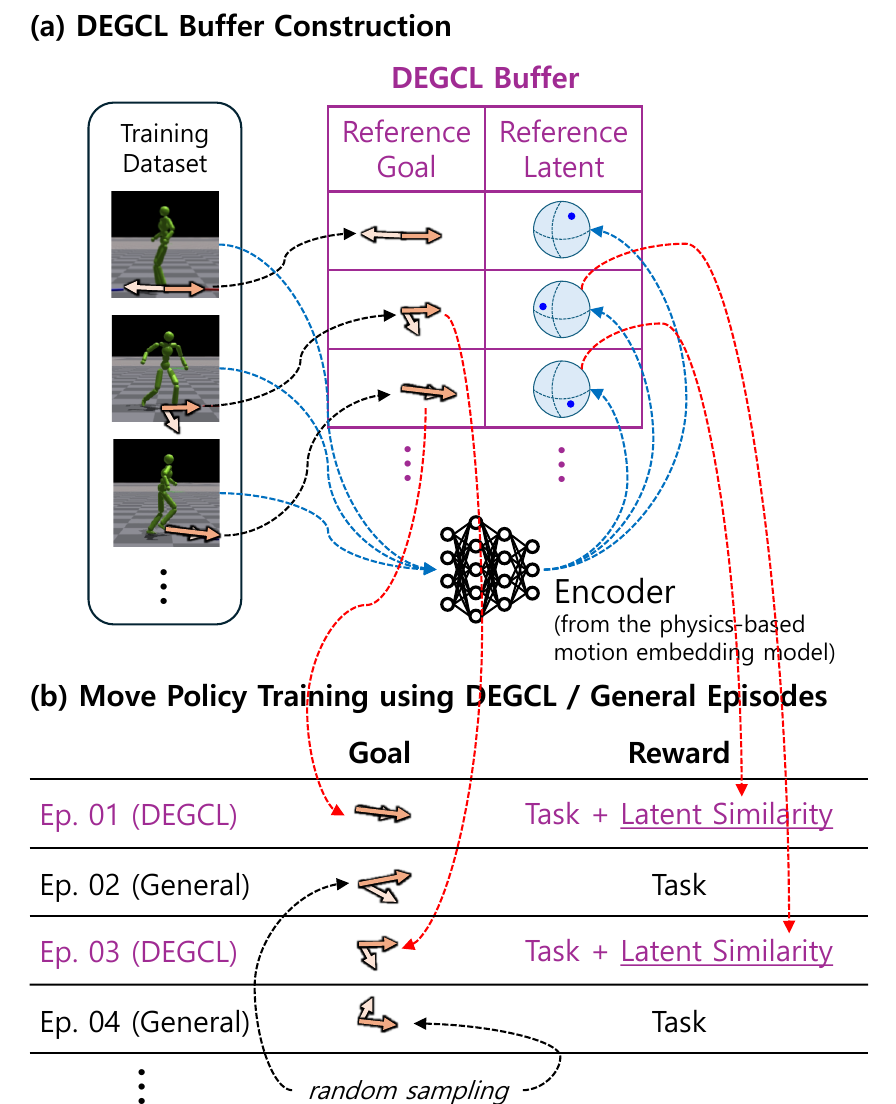}
  \caption{
  DEGCL process. 
  (a) Reference (goal, latent) pairs are extracted from the selected motion clips in the training dataset and stored in the DEGCL buffer $\mathcal{D}$.
  (b) During training, the Move policy learns from both DEGCL episodes and general episodes, allowing it to learn to perform goal-aligned actions for a variety of goals.
  }
\label{fig:degcl}
\end{figure}

To address this issue, we propose Data-Embedded Goal-Conditioned Latent Guidance (Figure~\ref{fig:degcl}).
This method leverages the goal-action relationships embedded in the motion data to guide the policy's latent output toward goal-aligned actions.
It consists of the following components:

\begin{description}

\item[DEGCL Buffer]
For selected motion clips from the training dataset, reference (goal, latent) pairs are stored in the buffer $\mathcal D$.
The horizontal movement velocity of the character’s root and its forward-facing direction on the horizontal plane are saved as the reference goal.
The reference latent is computed using the encoder of the pre-trained physics-based motion embedding model, CALM \cite{CALM} (Appendix~\ref{sec:appd-calm}) and stored in the buffer.

\item[DEGCL Episode]
In a proportion $p$ of the total training episodes, the DEGCL buffer is referenced to set the goal inputs and calculate the rewards.

\item[Goal Sampling for DEGCL Episodes]
In DEGCL episodes, the reference goal from a randomly selected pair in $\mathcal D$ is used as the input to the policy.
In contrast, general episodes use a randomly sampled goal as described earlier.

\item[Reward for DEGCL Episodes]
In DEGCL episodes, the reward is calculated as the sum of the task reward, which encourages goal achievement, and the latent similarity reward, which guides the policy to output a latent similar to the reference latent from the selected pair.
In contrast, general episodes calculate the reward using only the task reward.
Note that regardless of the episode type, the policy's input remains the same, so the reference latent is not included as an input to the policy.

\end{description}

The reason for utilizing not only DEGCL episodes but also general episodes for training is that the training motion dataset does not contain a sufficient variety of motions to evenly cover the full range of required goal inputs.
By experiencing randomly selected goal inputs in general episodes, the policy also learns how to achieve various goals that lie between the limited reference goals.
While the reward in general episodes does not directly enforce similarity between the reference latent and the output latent, the patterns learned from DEGCL episodes influence general episodes as well.
Consequently, the latent output for randomly selected goals in general episodes is indirectly influenced by the reference latent, enabling the trained policy to naturally perform actions observed in the motion dataset, even for goals that are not explicitly present in the dataset.

CALM~\cite{CALM} proposed precision training, which uses a latent similarity reward to train a policy to achieve various goals with a specific  motion (e.g., achieving different running directions with a "run" motion).
Our DEGCL extends this concept by expanding the learning scope beyond achieving diverse goals with a particular motion.
Instead, it leverages the information already embedded in the motion dataset about which motions are used to achieve specific goals.
This enables the policy to learn not only how to achieve various goals but also which motion to use for each goal, depending on the context (e.g., achieving a forward walking goal with a "normal walk" motion and a sideways walking goal with a "sideway walk" motion).

\paragraph{Reward}
The task reward for the Move policy is defined as follows:

\begin{equation}
\begin{aligned}
r^{\mathrm{mv\_task}}_t = 
0.7 &\ r^{\mathrm{vel}}_t \left( \mathbf v^{\mathrm{target}}_t, \mathbf v^{\mathrm{root(2)}}_t \right) + 
0.3 \ r^{\mathrm{dir}}_t \left( \mathbf d^{\mathrm{target}}_t, \mathbf d^{\mathrm{root}}_t \right), \\
\text{where} \quad 
&\mathbf v^{\mathrm{target}}_t = 
\begin{cases} 
\bar{\mathbf v}^{\mathrm{move}}_t, & \text{for DEGCL episodes} \\ 
\hat{\mathbf v}^{\mathrm{move}}_t, & \text{otherwise}
\end{cases} \\
\text{and} \quad 
&\mathbf d^{\mathrm{target}}_t = 
\begin{cases} 
\bar{\mathbf d}^{\mathrm{face}}_t, & \text{for DEGCL episodes} \\ 
\hat{\mathbf d}^{\mathrm{face}}_t, & \text{otherwise}
\end{cases}
\label{eq:r_move_task}
\end{aligned}
\end{equation}
where $\mathbf d^{\mathrm{root}}_t$ represents the unit vector of the character root's facing direction on the horizontal plane, expressed in the global coordinate system.
$\hat{\mathbf v}^{\mathrm{move}}_t$ and $\hat{\mathbf d}^{\mathrm{face}}_t$ denote the randomly sampled goal inputs used in general episodes, while $\bar{\mathbf v}^{\mathrm{move}}_t$ and $\bar{\mathbf d}^{\mathrm{face}}_t$ represent the reference goal inputs selected from $\mathcal D$ in DEGCL episodes.
$r^{\mathrm{vel}}_t(\cdot)$ encourages the horizontal velocity of the character root to match the target movement velocity and, similar to Equation~\ref{eq:r_drib}, includes normalization by the target speed. Meanwhile, $r^{\mathrm{dir}}_t(\cdot)$ encourages the facing direction of the character root to align with the target facing direction.
For a detailed explanation of each term, please refer to Appendix~\ref{sec:appd-move-task-reward}.

The latent similarity reward is defined as follows:
\begin{equation}
r^{\mathrm{lt\_sim}}_t = \bar{\mathbf z}_t \cdot \mathbf z_t,
\label{eq:r_move_ltsim}
\end{equation}
where $\bar{\mathbf z}_t$ denotes the reference latent selected from $\mathcal{D}$, and $\mathbf z_t$ represents the latent output by the policy.
Considering the characteristics of CALM~\cite{CALM}, the physics-based motion embedding model we used, which represents each motion embedding projected onto an $l_2$ unit hypersphere, we employed cosine similarity.

The overall reward for the Move policy is defined as follows:
\begin{equation}
r^{\mathrm{move}}_t = 
\begin{cases} 
0.5 \ r^{\mathrm{mv\_task}}_t + 0.5 \ r^{\mathrm{lt\_sim}}_t, & \text{for DEGCL episodes} \\ 
r^{\mathrm{mv\_task}}_t. & \text{otherwise}
\end{cases}    
\label{eq:r_move}
\end{equation}

\paragraph{Episode Initialization}
Unlike other policies, episodes for training the Move policy are initialized not with STI but with the character in a rest pose, standing upright and facing forward.
This is because, unlike other policies that need to handle the ball skillfully during transitions, the Move policy does not involve interacting with the ball, making it unnecessary for episodes to begin from the state of a previous skill.
In practice, initializing episodes with the character always in the rest pose still resulted in effective learning.

\paragraph{Episode Termination}
Each episode terminates after 10 seconds.

\subsection{Kick}
Kicking is a technique that uses the foot to strike the ball and is used for actions such as shooting, passing, and clearing. 
Our Kick policy is trained to enable the ball to be kicked at a user-specified desired velocity.

\paragraph{Goal Input}

The Kick policy takes the target kick velocity $\hat{\mathbf v}^{\mathrm{kick}}_t \in \mathbb{R}^3$ as its goal.
During training, the horizontal direction is uniformly sampled from [-45, $45^\circ$] relative to the character's forward direction, the vertical direction from [0, $45^\circ$], and the speed from [5, \SI{35}{m/s}].
The actual input to the policy is the target velocity $\hat{\mathbf v}^{\mathrm{kick}}_{t\{c\}}$, expressed in the character coordinate system.

\paragraph{Reward}
The reward for the Kick policy is defined as follows:

\begin{equation}
r_t^{\mathrm{kick}} = \mathrm{exp} \left( - \left(  \frac{\left\| \hat{\mathbf v}^{\mathrm{kick}}_t - \mathbf v^{\mathrm{ball(3)}}_t \right\|} {\|\hat{\mathbf v}^{\mathrm{kick}}_t\| +  \epsilon} \right)^2 \right).
\label{eq:r_kick}
\end{equation}

Equation~\ref{eq:r_kick} is evaluated for only a very short duration (1/3 second) after the collision, guiding the policy to ensure that the initial velocity of the kicked ball closely matches the target velocity $\hat{\mathbf v}^{\mathrm{kick}}_t$.
Similar to Equation~\ref{eq:r_drib}, it incorporates normalization by the target speed.

\paragraph{Episode Initialization}

In our PhysicsFC FSM, the Kick policy transitions from the Dribble policy (Figure~\ref{fig:fsm}). However, initializing episodes solely from the Dribble STI buffer does not sufficiently train the ability to kick the ball from a stationary or slow-moving state.
To address this, 70\% of episodes are initialized from the Dribble STI buffer, while the remaining 30\% are initialized with the character in a rest pose and the ball placed at a random position within a radius of \SI{2}{m} around the character's root.

\paragraph{Episode Termination}
An episode ends 1/3 second after the ball collides with the character.
If no collision occurs within 3 seconds, the episode is terminated.

\section{PhysicsFC FSM for Runtime Control}\label{sec:fsm}

To enable a seamless sequence of physics-based football plays during runtime driven by user intent,
such as progressing from ball possession through dribbling and then to passing or shooting,
we propose a PhysicsFC FSM (Figure~\ref{fig:fsm}).
Each learned skill policy is defined as a state within the FSM, and transitions between skill policies are governed by transitions defined within the FSM.  

Each transition condition is based on user input or surrounding context, depending on the skill before and after the transition, as follows:
\begin{description}

\item[Move $\rightarrow$ Trap] The transition occurs when both the user inputs the command to start trapping and the ball is approaching.
Thus, while the ball is in the air, the user can use the Move policy to position the character and initiate the trap command at the desired moment, transferring control to the Trap policy.

\item[Move $\rightarrow$ Dribble] 
This occurs when two conditions are met: the horizontal distance between the ball and the character’s root is within 2 meter, and the ball is moving toward the character (i.e., the distance between them is decreasing).
The requirement for the ball to be approaching was added to prevent an immediate transition to Dribble right after switching from Kick to Move.
As a result, when the user moves the character without possession toward the ball, the character naturally starts dribbling as it gets closer to the ball.

\item[Dribble $\rightarrow$ Kick] It occurs when the user inputs the command to start kicking.

\item[Dribble $\rightarrow$ Move] It occurs when the horizontal distance between the ball and the character exceeds 2 meters.

\item[Trap $\rightarrow$ Dribble] It happens when the ball and the character collide.
Thus, the character immediately begins dribbling right after making the first touch to trap the ball.

\item[Trap $\rightarrow$ Move] It happens either when the user inputs the command to end trapping or the ball is moving away.
Thus, at any point while the character is moving under Trap policy, the user can cancel trapping and regain direct control of the character’s movement through Move policy.

\item[Kick $\rightarrow$ Move]
This happens either when the ball and the character collide or when the horizontal distance between them exceeds 2 meters.
The distance condition was introduced to let the user regain direct control of the character’s movement if the ball unexpectedly moves away during a kick attempt (e.g., when trying to kick a ball that moves too quickly past the character).

\item[Kick $\rightarrow$ Dribble] It occurs when the user inputs the command to end kicking. 
Thus, at any point while the character is preparing to kick, the user can cancel the kick and resume dribbling.

\end{description}

\section{Implementation and Training}

\subsection{Training Details}
\label{sec:training-details}

\paragraph{Motion Dataset}

We trained the low-level policy using a commercially available football motion dataset.
The dataset includes approximately three minutes of motion data composed of 90 clips, featuring various football movements such as locomotion, jumping, dribbling, kicking, and passing.
For more details on the motion dataset, please see Appendix~\ref{sec:appd-dataset}.

\paragraph{STI Buffer Construction}

The Move STI buffer is constructed by collecting 50,000 character states sampled at random time points during simulations using the trained Move policy.
Random goal inputs, sampled in the same manner as during Move policy training, are provided as inputs to the policy.

The Trap STI buffer is created by collecting 50,000 states of the character and the ball at collisions.
These states are generated using the trained Trap policy, with initial character and ball states, as well as target body inputs, sampled in the same way as during training.

The Dribble STI buffer is created by collecting 50,000 states of the character and the ball sampled at random time points during simulations.
These simulations use the trained Dribble policy, with initial character and ball states and goal inputs sampled in the same manner as during training.

No skill policy uses the kick states for training, so the Kick STI buffer is not created.

\paragraph{Skill Policy Training Procedure Considering STI}

We trained these policies in the following sequential order for the STI process, which uses simulated states of other skill policies for episode initialization to enable smooth transitions between skill policies:
Move, Trap (using the Move STI buffer), Dribble (using the Move and Trap STI buffers), and Kick (using the Move and Dribble STI buffers).

When using STI, the sequential dependency between policies naturally defines a training order, which could theoretically create a cycle in the training process. 
However, even if a cycle arises, it is practically feasible to train one policy without STI first and then fine-tune it later using STI, thereby allowing all policies in the cycle to be trained with STI.
In our case, there was no policy that relied on simulated states from the trained Kick policy for STI. Therefore, it was unnecessary to employ this approach.

\paragraph{DEGCL Configuration}

The DEGCL buffer was constructed from 16 motion clips selected from a total of 90 motion clips used during training. 
These clips involve motions that maintain various frontal orientations while moving, including forward walk/jog/run, backward walk/jog, lateral walk, forward diagonal ($45^\circ$) jog, backward diagonal ($45^\circ$) jog motions, and their mirrored versions.

Since the movement and frontal directions remain consistent within each motion clip, we used the average movement direction and frontal direction of each clip as the reference goal, denoted as $\bar{\mathbf v}^{\mathrm{move}}$ and $\bar{\mathbf d}^{\mathrm{face}}$, respectively. 
The physics-based motion embedding model we used, CALM~\cite{CALM}, maps short motion clips of approximately 2 seconds into a single latent vector.
Using this model, each selected motion clip, ranging in length from 0.5 to 1.5 seconds, was mapped to a latent vector, which was then used as the reference latent $\bar{\mathbf z}$.

For training the Move policy, 80\% of the episodes were DEGCL episodes ($p=0.8$), while the remaining 20\% were general episodes.

\paragraph{Trap Policy Training}

The training of the Trap policy is conducted in two stages:
In the first stage, episodes are initialized only with lob passes to train the policy to control incoming airborne balls.
In the second stage, episodes are initialized with an 80\% probability of a lob pass and a 20\% probability of a ground pass, allowing the policy to handle both types of passes.
For lob passes, the initial ball position $\mathbf p_\text 0$ is analytically calculated to ensure the ball lands at $\mathbf p_\text f$.
However, due to numerical integration during state updates in the physics simulation, slight deviations from the intended landing position may occur.
The policy is trained to account for these deviations and successfully trap the ball despite such discrepancies.

\paragraph{Recovery from Falls}

To enable characters to recover from falls, we adopt the recovery strategy used by ASE \cite{ASE} and CALM \cite{CALM} for training our low-level policy.
During training, the character is initialized in random fallen states with a 10\% probability, allowing it to learn effective recovery strategies.
As a result, the low-level policy enables the character to automatically recover and seamlessly continue performing football skills after losing balance.

\subsection{Training Environment}

\paragraph{Hardware and Training Time}

Both the low-level policy and the skill policies were trained using PPO on a single NVIDIA A6000 GPU.
The low-level policy was trained with 2,048 environments over 30 days, processing a total of 5 billion samples.
The Move policy was trained with 4,096 environments for 44 hours, using 6 billion samples.
The Trap policy was trained with 4,096 environments for 24 hours, processing 3.2 billion samples.
The Dribble policy was trained with 4,096 environments for 40 hours, using 6 billion samples.
Finally, the Kick policy was trained with 4,096 environments for 30 hours, processing 5.5 billion samples.
For details on the \textcolor{rv}{network architecture and training time}, please refer to the Appendix~\ref{sec:appd-network}.

\paragraph{Simulation}

We used Isaac Gym as the physics simulation engine.
The simulation frequency is \SI{60}{Hz}, and both the low-level policy and skill policies are executed at \SI{30}{Hz}.
To achieve natural ball control, the physical properties of the ball were configured to closely match real-world values.
The ball's diameter is set to \SI{22}{cm}, consistent with the size used in official matches, its mass is set to \SI{450}{g}, and the coefficient of restitution is set to 0.8.
For more details on the physics simulation configuration, please refer to Appendix~\ref{sec:appd-simulation}.

\begin{figure}
  \centering
  \subfigure[Dribble]{
  \includegraphics[trim=50 0 0 0, clip, width=.475\linewidth]{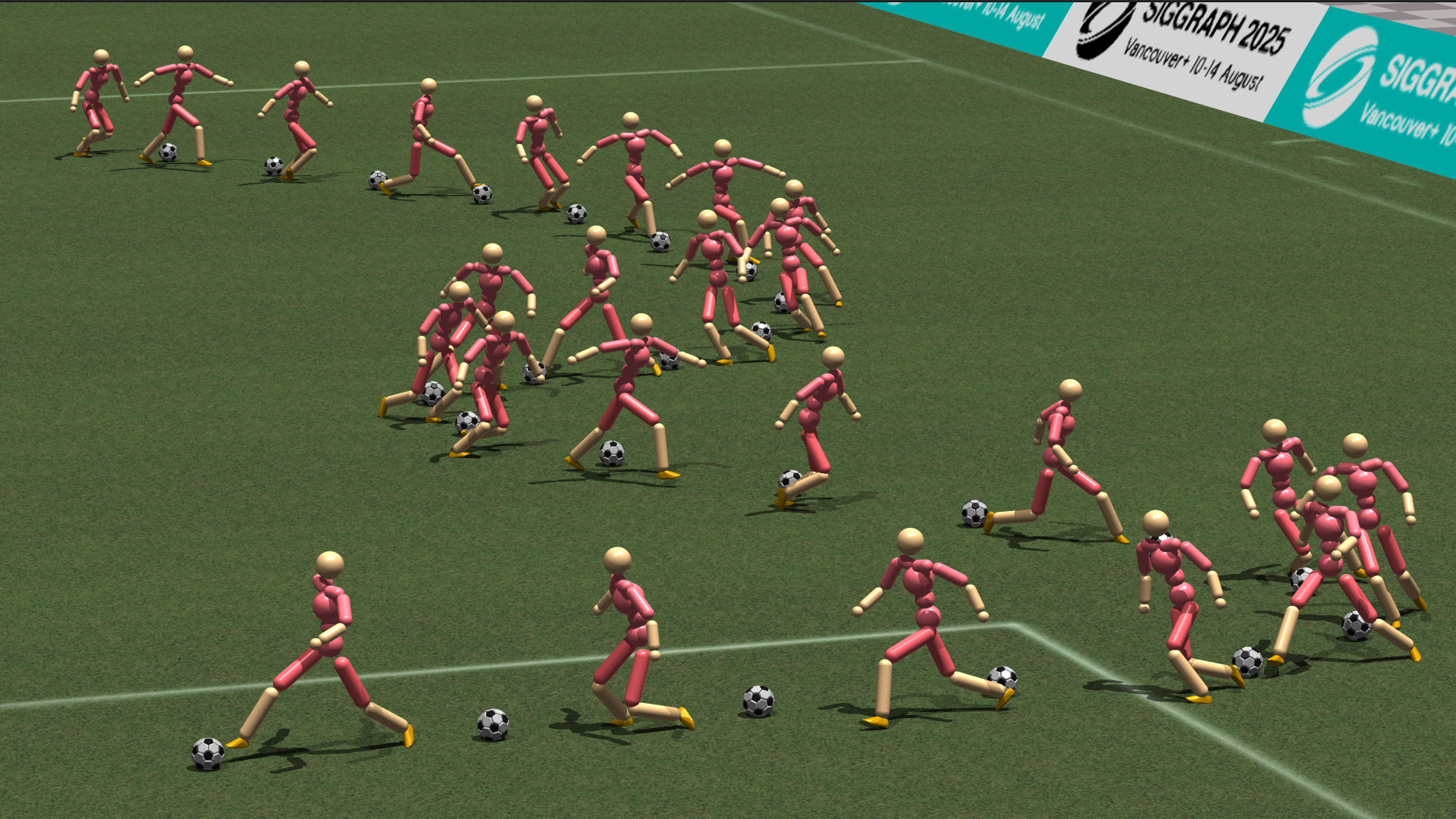}
  }
  \subfigure[Trap]{
  \includegraphics[trim=200 157 200 40, clip, width=.475\linewidth]{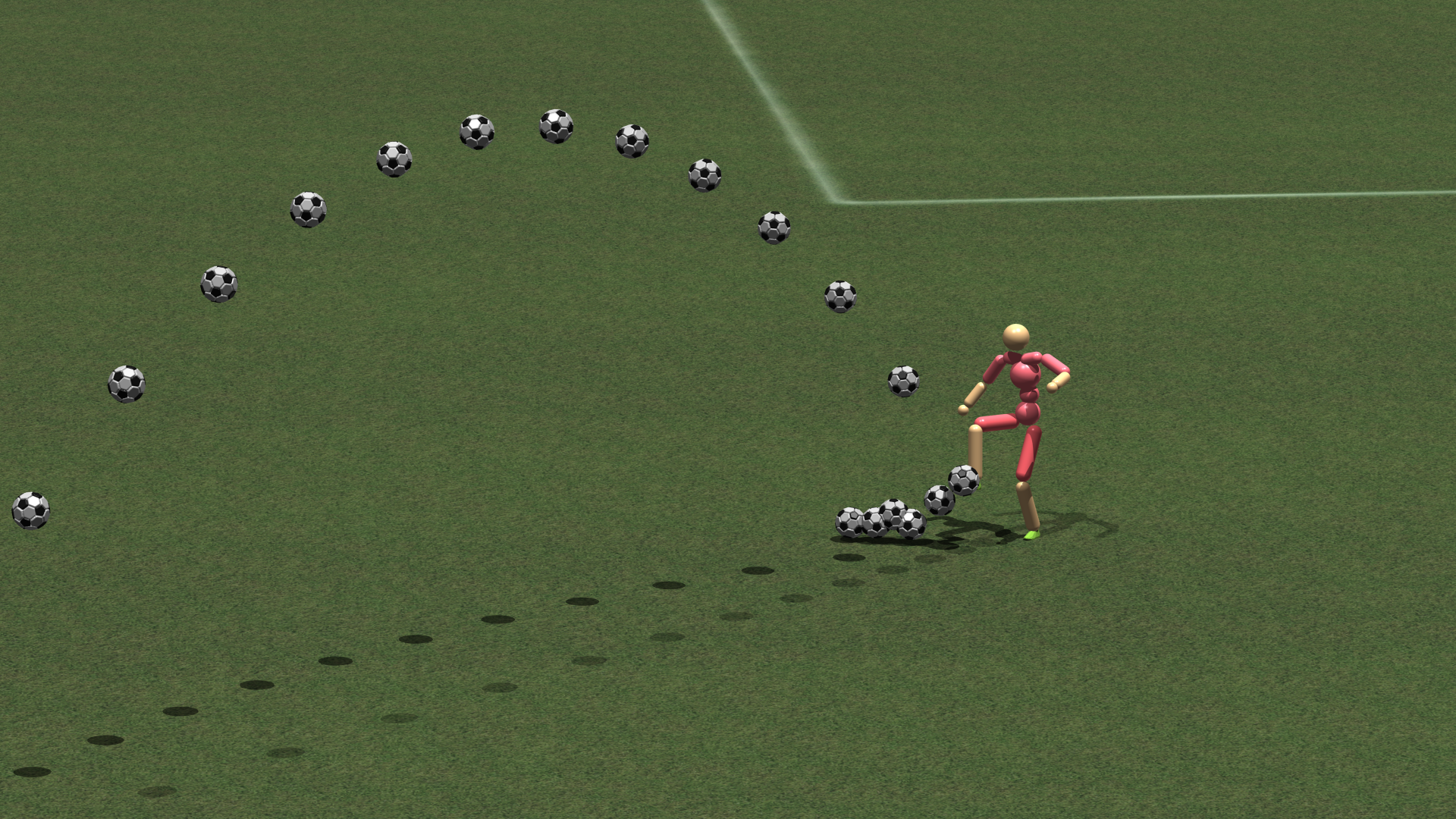}
  }
  \subfigure[Kick]{
  \includegraphics[trim=200 166 300 90, clip, width=.475\linewidth]{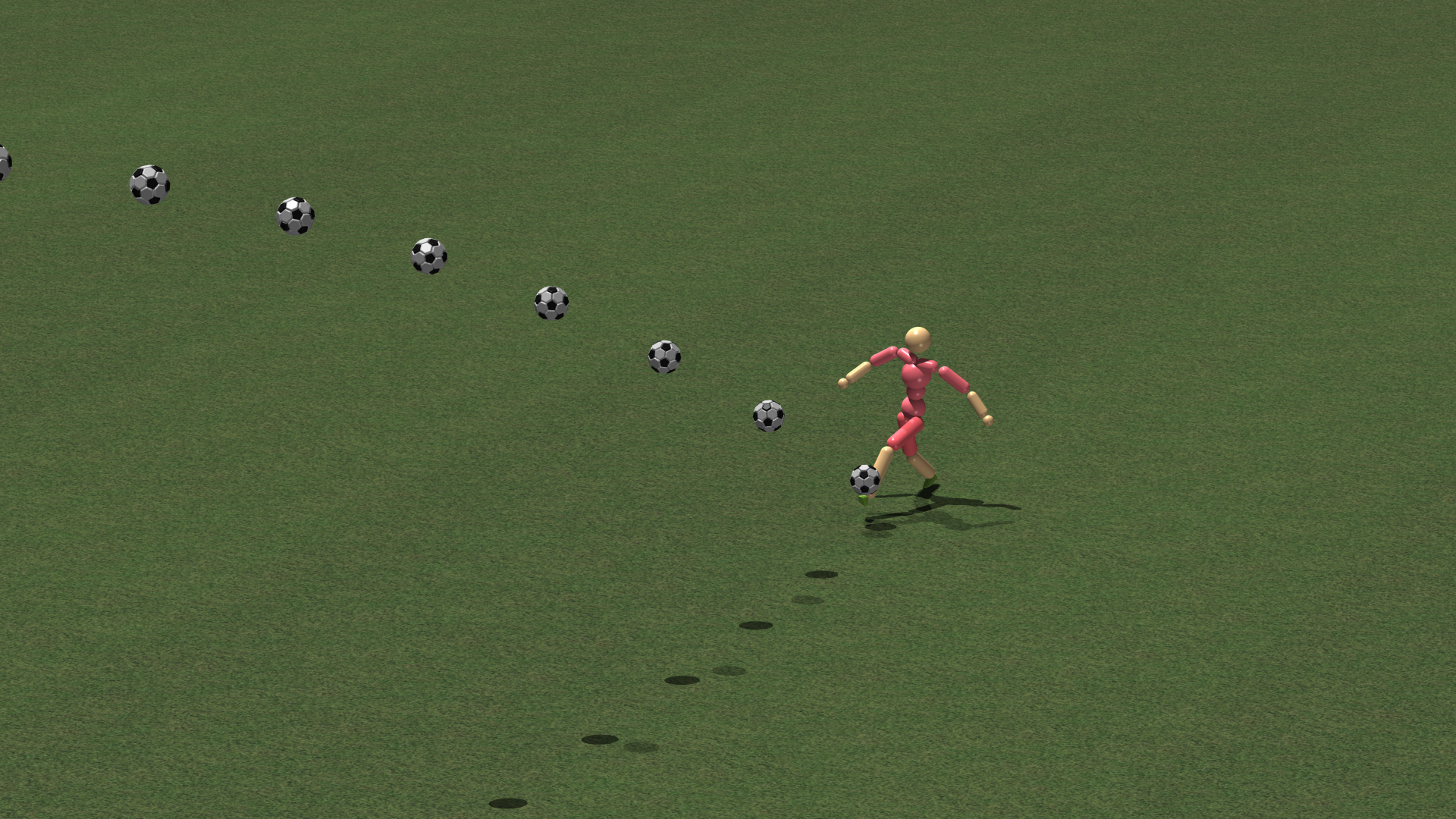}
  }
  \subfigure[Move]{
  \includegraphics[trim=0 0 30 0, clip, width=.475\linewidth]{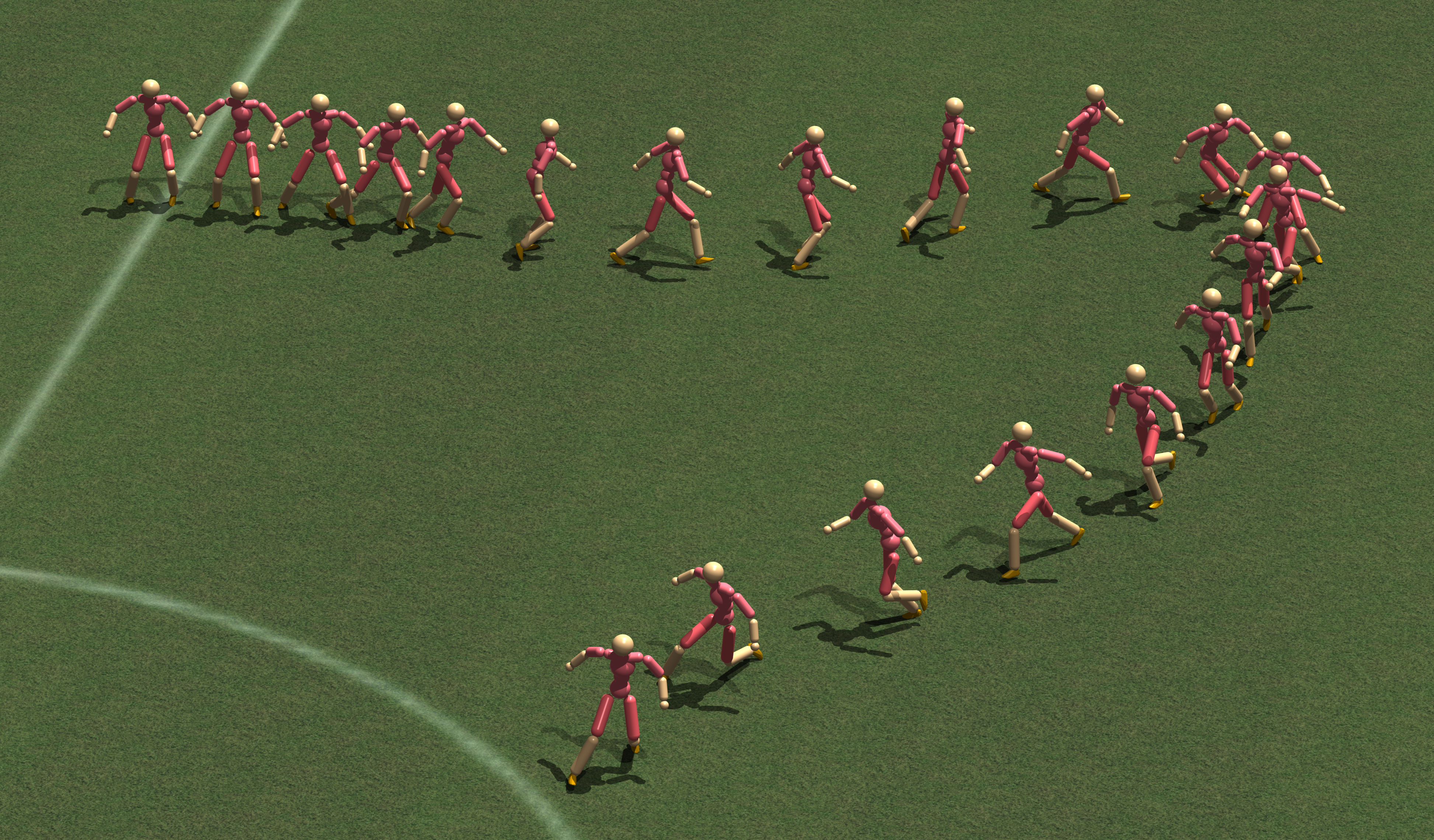}
  }
  \caption{Football skills performed by PhysicsFC.}
\label{fig:skills}
\end{figure}

\section{Interactive Demos}
\label{sec:interactive-demo}

We present several interactive demos as examples of potential applications of the PhysicsFC, demonstrating its effective applicability to various interactive scenarios.
In these demos, all characters (including the controlled player, teammate players, and opponent players) are driven by the PhysicFC FSM (Figure~\ref{fig:fsm}).
This section provides an overview of each demo.
Details regarding the specific implementation of each demo can be found in the Appendix~\ref{sec:appd-interactive-demo}.
For details on the gamepad input used at runtime, please see the Appendix~\ref{sec:appd-gamepad}.

\paragraph{User-Controlled Football Player}

The user can control a single football player character to perform football skills such as moving, dribbling, trapping, and kicking (Figure~\ref{fig:skills}) in response to various situations and seamlessly transition between these skills using the proposed PhysicsFC FSM.
This enables the user to control the character to trap the ball while moving, transition to dribbling, and then perform a pass or a shot as part of typical football gameplay (Figure~\ref{fig:teaser}).

\paragraph{User-Controlled Give and Go Play}

\begin{figure}
  \centering
  \includegraphics[trim=240 120 240 120, clip, width=1.\linewidth]{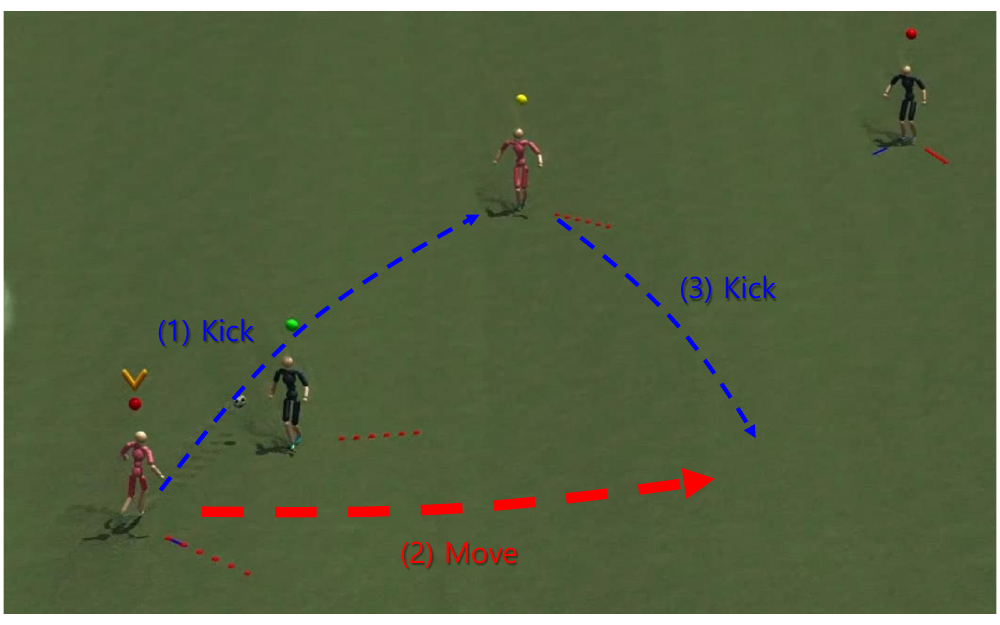}
  \caption{User-controlled give and go play.
  In this and subsequent figures, the v mark above a character's head indicates the player controlled by the user, while the color of the sphere above the head represents the policy currently governing each character (Red: Move, Yellow: Trap, Green: Dribble, Blue: Kick).
  The sequence of red spheres on the ground represents the target dribble or movement velocity, while the blue bar indicates the target forward-facing direction.}
\label{fig:give-and-go}
\end{figure}

This demo presents a scenario where the controlled player, guided by the user’s gamepad inputs, performs various football skills while collaborating with a teammate.
The user controls the controlled player to pass the ball to a teammate, who then returns the ball, enabling seamless team play (Figure~\ref{fig:give-and-go}).
Throughout this process, each player transitions between appropriate policies via the PhysicsFC FSM, ensuring smooth and context-aware actions based on user inputs and game situations.

\paragraph{Competitive Trapping and Dribbling}

\begin{figure}
  \centering
  \includegraphics[trim=60 0 60 0, clip, width=1.\linewidth]{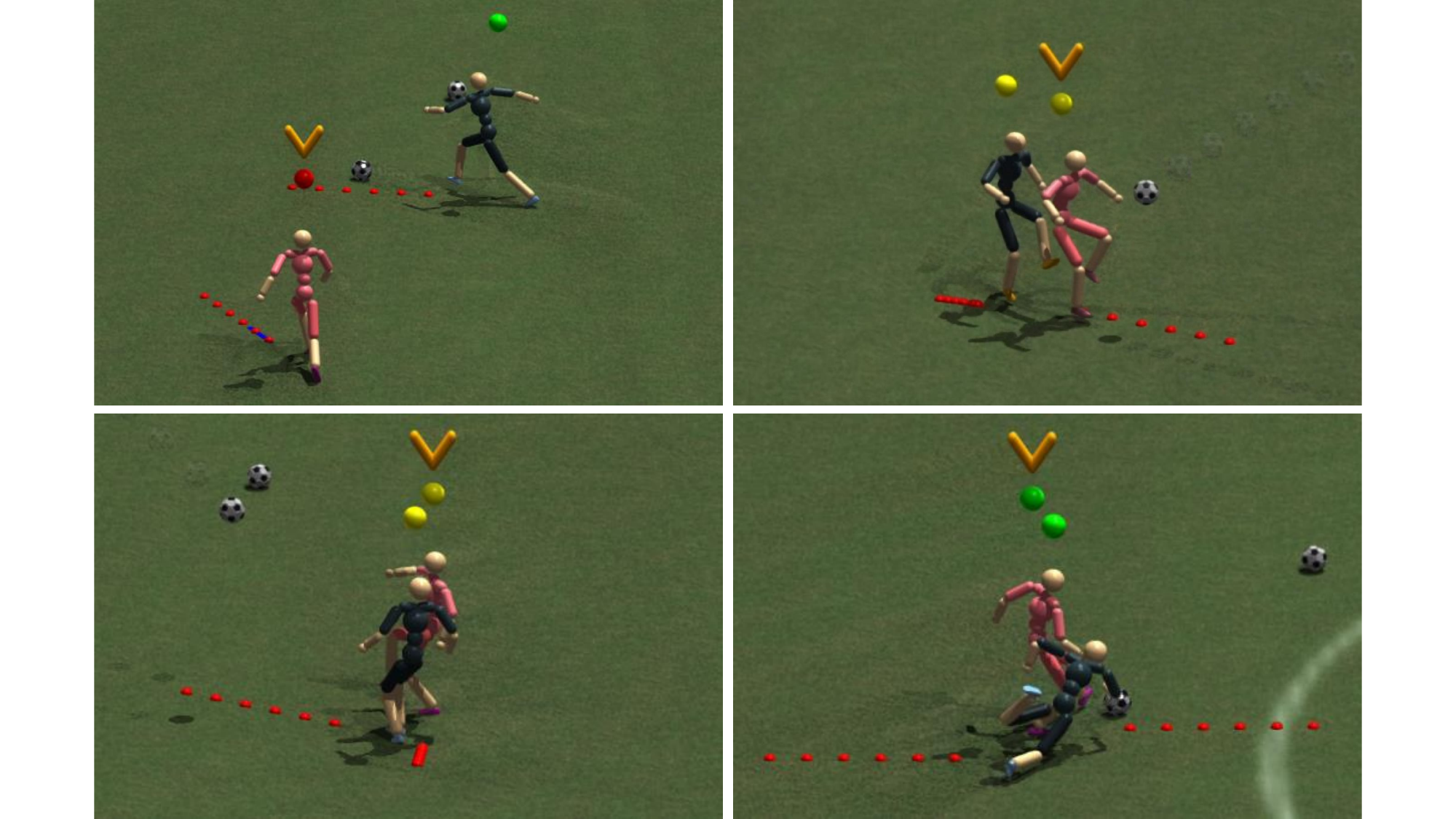}
  \caption{Competitive trapping and dribbling.
  }
\label{fig:competitive}
\end{figure}

This demo features a competitive scenario where the controlled player and an opposing player repeatedly attempt to trap and dribble the ball, competing for possession (Figure~\ref{fig:competitive}). It highlights the strengths of PhysicsFC, as both the characters and the ball are fully governed by physics simulation.
Each skill policy is trained to enable characters to recover and stand up after falling (Section~\ref{sec:training-details}), allowing them to continue performing the appropriate skills during gameplay. Despite being trained without explicitly accounting for opposing players, our skill policies demonstrate the ability to produce natural physical interactions between characters.

\paragraph{Simulated 11v11 Football Game with User-Controlled Player Switching}

\begin{figure}
  \centering
  \includegraphics[trim=0 200 500 50, clip, width=1.\linewidth]{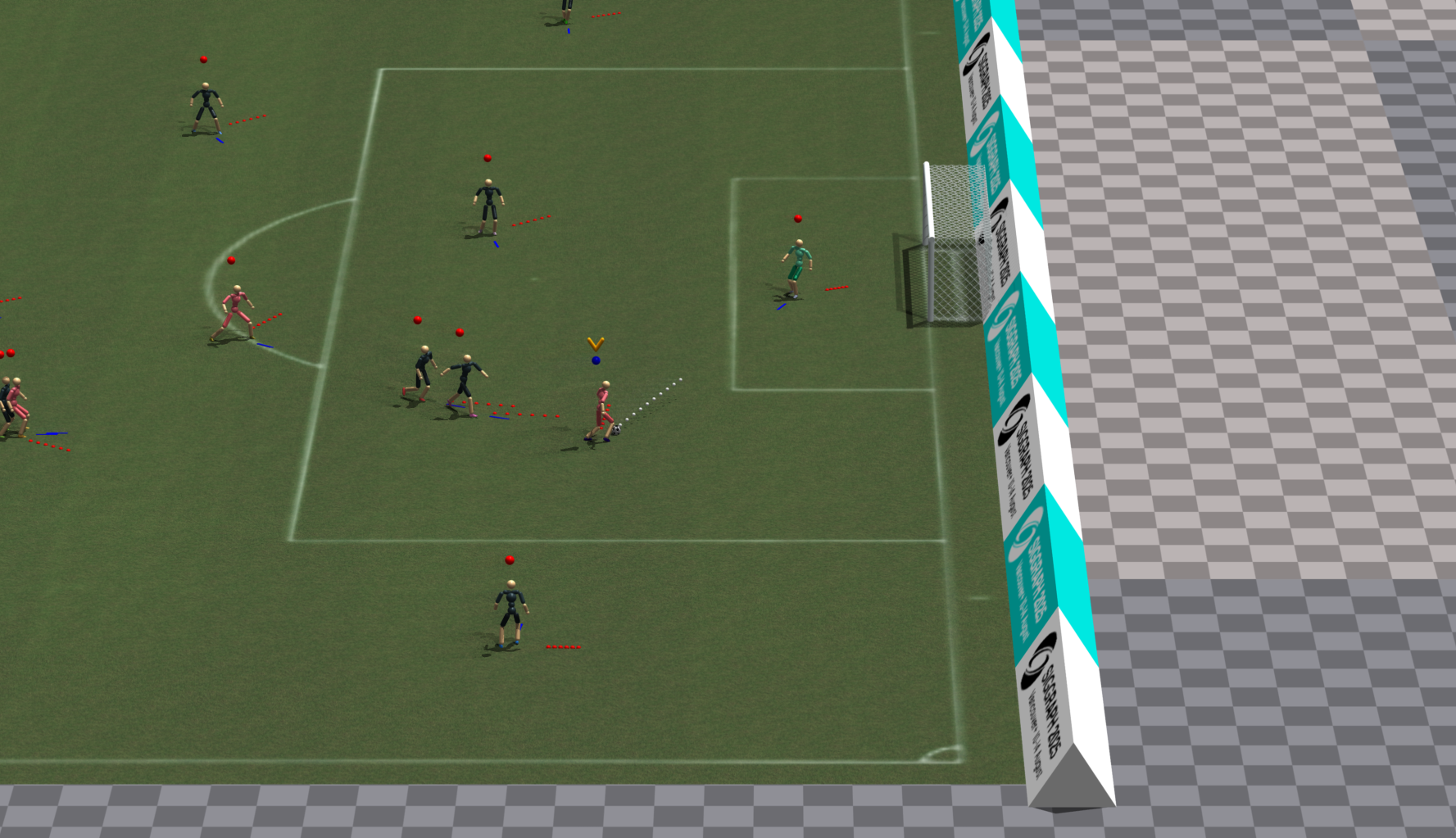}
  \caption{
  Simulated 11v11 football game with user-controlled player switching.
  }
\label{fig:11v11}
\end{figure}

This demo features a full-scale football game scenario with 11 players on the user’s team and 11 players on the opposing team, for a total of 22 PhysicsFC agents moving simultaneously in a physics simulation environment.
The user can switch the controlled player in real time using the gamepad to perform actions such as passing, dribbling, and kicking to influence the match.
Control is automatically transferred after a pass.
Each player’s Move policy input is adjusted to maintain their positions in a 4-3-1-2 formation, which dynamically shifts based on the ball’s location.
This demo highlights the scalability of PhysicsFC and its ability to handle complex multi-agent environments with realistic, physics-based interactions.

\section{Skill Policy Evaluation}
\label{sec:skill-eval}

To validate the effectiveness of our football skill policy training method, we conducted quantitative performance comparisons between each football skill policy and its ablated models.

\subsection{Dribble}

We conducted the following two experiments to evaluate the performance of the Dribble policy.

The first experiment aimed to evaluate performance in various dribbling scenarios by measuring metrics for 1,000 randomly and repeatedly generated goals during a continuous dribbling process.
The target dribble velocity was reset 1,004 times at 5-second intervals, with directions sampled uniformly from all directions and speeds within the range [1.0, \SI{7.0}{m/s}].
To account for potential instability during the initial ball control phase, measurements began from the fourth goal reset. All comparison models were evaluated using the same test case, consisting of $1004 \times 5$ seconds. The simulation started with the character in a default standing pose, and the ball was positioned \SI{1}{m} away in a random direction around the character.

We defined the following metrics and conducted measurements based on them:

\begin{description}
\item[Character-Ball Distance (CBD)]
The average horizontal difference between the character's root (pelvis) position and the ball's position, calculated across all frames.

\item[Foot-Ball Distance (FBD)]
The average distance between the horizontal position of the foot at the moment of ground contact and the ball's position at that time, calculated across all contacts.

\item[Dribbling Goal Achievement Rate (DGAR)]
Once a goal is set, it is considered achieved if the magnitude of (ball velocity - target velocity) comes within 10\% of the target speed before the next goal reset (within 5 seconds).
The metric is defined as the number of achieved goals divided by the number of evaluated goals.

\end{description}

We measured the metrics for Ours and the following ablated models:
\begin{description}
    \item[Dribble-w/o $r^{\mathrm{ball\_vel}}_t$, $r^{\mathrm{ball\_root\_pos}}_t$, $r^{\mathrm{root\_vel}}_t$]
    The Dribble policy learned excluding each reward term in Equation~\ref{eq:r_drib}.
    
    \item[Dribble-w/o DistanceET] 
    The Dribble policy learned without early termination, where the horizontal distance between the ball and the character's root exceeds \SI{3}{m}.
    
    \item[Dribble-w/o NTS] 
    The Dribble policy learned without normalization by target speed.
    
    \item[Dribble-w/o BoxFoot] 
    The Dribble policy learned by replacing the character's foot with a simple box-shaped foot instead of a football boots mesh.
    
\end{description}

\begin{table}[h]
    \centering
    \caption{Comparison of our Dribble policy and its ablated models.}
    \begin{tabular}{lccc}
        \toprule
        & CBD(m)$\downarrow$ & FBD(m)$\downarrow$ & DGAR(\%)$\uparrow$ \\
        \midrule
        Dribble-w/o $r^{\mathrm{ball\_vel}}_t$ &  1.05 & 0.86 & 5.9 \\
        Dribble-w/o $r^{\mathrm{ball\_root\_pos}}_t$ & 0.86 & 0.75 & 75.4 \\
        Dribble-w/o $r^{\mathrm{root\_vel}}_t$ & 0.91 & 0.81 & 73.5 \\
        Dribble-w/o DistanceET & 9.06 & 8.72 & 2.0 \\
        Dribble-w/ BoxFoot & 1.05 & 0.87 & 21.6 \\
        Dribble-w/o NTS & 0.90 & 0.75 & 71.7 \\
        Dribble-Ours & \textbf{0.77} & \textbf{0.60} & \textbf{90.3} \\
        \bottomrule
    \end{tabular}
    \label{tbl:dribble-1}
\end{table}

As shown in Table~\ref{tbl:dribble-1}, models trained without any of the three reward terms performed worse across all metrics compared to Ours.
Specifically, the w/o $r^{\mathrm{ball\_vel}}_t$ model almost failed to achieve the target dribble velocity.
The w/o DistanceET model showed the most significant performance degradation, failing to learn how to dribble at all.
The w/ BoxFoot model learned how to dribble, but frequently failed to change direction, achieving a much lower target achievement rate.
The w/o NTS model was able to learn how to dribble with direction changes, but its transitions were not as swift as those in Ours, and it underperformed across all metrics compared to Ours.
Ours achieved the shortest ball-character and ball-foot distances and recorded the highest target achievement rate.

In the second experiment, the performance of the Dribble policy was evaluated by measuring the proposed quantitative metrics for different target speeds.
At the start of the simulation, the character is in a default standing pose with the ball positioned 1 meter in front of them.
After the simulation begins, the direction of the target dribble velocity is set to the character's front, and the magnitude is set to the respective target speed, continuously provided as the goal input for the policy.
To ensure that the character and ball's movement has stabilized, measurements were taken starting 10 seconds after the simulation began and continued for 30 seconds. In this experiment, the following additional metrics were measured:
\begin{description}
    \item[Chracter Speed (CS)]
    The average horizontal speed of the character's root.
\end{description}

\begin{table}[h]
    \centering
    \caption{Dribble policy performance evaluation by target speed.}

    \begin{tabular}{lcccc}
        \toprule
        & CS(m/s) & CBD(m) & FBD(m) \\
        \midrule
        Dribble-Ours (1m/s) & 1.07 & 0.59 & 0.45 \\
        Dribble-Ours (2m/s) & 2.07 & 0.74 & 0.55 \\
        Dribble-Ours (3m/s) & 3.08 & 0.83 & 0.61 \\
        Dribble-Ours (4m/s) & 4.04 & 0.99 & 0.78 \\
        Dribble-Ours (5m/s) & 4.99 & 0.95 & 0.68 \\
        Dribble-Ours (6m/s) & 5.56 & 1.04 & 0.76 \\
        Dribble-Ours (7m/s) & 5.73 & 1.11 & 0.80 \\
        
        \bottomrule
    \end{tabular}
    \label{tbl:dribble-2}
\end{table}

As shown in Table~\ref{tbl:dribble-2}, our Dribble policy demonstrated an actual character movement speed that closely matched the target dribble speed up to \SI{5}{m/s}, but started to fall short from \SI{6}{m/s} onward. Both FBD and CBD were found to increase as the target speed increased. Generally, dribbling while running fast is considered more difficult than dribbling at slower speeds, and this was similarly observed in our Dribble policy.

\subsection{Trap}

To evaluate the performance of the Trap policy, we conducted an experiment where we measured metrics while trapping 1,000 lob passes with various trajectories. The target body part for the policy's input was randomly selected from six body parts (head, torso, either lower leg, or either foot). The initial state of the ball was set using the same random sampling method for lob passes used during Trap policy training, and the character's initial state was set to a default standing pose. All comparison models were evaluated using the same test case, consisting of 1,000 lob passes generated in this manner.

We defined the following metric to evaluate the Trap policy:
\begin{description}
\item[Trapping Success Rate (TSR)]
The success rate of trapping lob passes. A trap is considered successful if the character touches the ball before it hits the ground.

\item[Handball Ratio in Trapping Success (HRTS)]
The ratio of successful traps where the ball touches the handling body parts to the total number of successful traps.

\item[Relative Ball Speed Post-Trap (RBSPT)]
The magnitude of the difference vector between the root velocity and the ball velocity, averaged over 5 frames immediately after ball contact.

\end{description}

We measured the metrics for Ours and the following ablated models:
\begin{description}
\item[Trap-w/o $r^{\mathrm{before}}_t$, $r^{\mathrm{after}}_t$]
The Trap policy learned without each reward terms in Equation~\ref{eq:r_trap}.

\item[Trap-w/o HandArmET]
The Trap policy learned without the early termination condition where the ball touches the character’s hand, forearm, or upper arm.

\item[Trap-w/o ProjectileInit]
The Trap policy learned by randomly sampling the ball's vertical launch angle, initial velocity, and landing position in the same way as during Trap policy training, but instead of analytically calculating the distance from the landing position to the initial position as described in Appendix~\ref{sec:appd-ball-distance-angle}, the distance was randomly sampled within the range of [10, \SI{20}{m}].

\end{description}

\begin{table}[h]
    \centering
    \caption{Comparison of our Trap policy and its ablated models.}
    \begin{tabular}{lcccc}
        \toprule
        & TSR(\%)$\uparrow$ & HRTS(\%)$\downarrow$ & RBSPT(m/s)$\downarrow$\\
        \midrule
        Trap-w/o $r^{\mathrm{before}}_t$ & 28.6 & \textbf{5.1} & 4.75 \\
        Trap-w/o $r^{\mathrm{after}}_t$ & 74.2 & 9.3 & 4.43 \\
        Trap-w/o ProjectileInit & 21.1 & 5.2 & 5.22 \\
        Trap-w/o HandArmET & 77.1 & 20.7 & 3.94 \\
        Trap-Ours & \textbf{78.3} & 5.6 & \textbf{3.69} \\
        \bottomrule
    \end{tabular}
    \label{tbl:trap}
\end{table}

As shown in Table~\ref{tbl:trap}, the w/o ProjectileInit model had the lowest success rate for touching the ball and the highest magnitude of the ball's relative velocity immediately after contact. This suggests that our proposed projectile dynamics-based ball-state initialization plays a crucial role in learning a successful trapping policy.
The w/o $r^{\mathrm{before}}_t$ model, lacking any mechanism to encourage the character to approach the ball during training, had the next lowest success rate.
Both of these models had a slightly lower handball foul rate compared to Ours, as they failed to learn how to approach the ball effectively. Even though they touched the ball, the contact was often minimal, especially with the feet. In contrast, Ours learned to approach the ball and receive it with the designated body part, which led to a slightly higher chance of the ball touching the hands or arms while attempting to move toward it.
The w/o HandArmET model had a significantly higher handball foul rate compared to the other models. Ours achieved the highest success rate for touching the ball, with the least ball bounce after contact and the lowest level of handball foul rate.

\subsection{Move}

To evaluate the performance of the Move policy, we measured metrics for 1,000 randomly and repeatedly generated goals during a continuous moving process.
The facing direction was randomly sampled from all directions at 5-second intervals, while the target movement velocity was randomly sampled within the range [1.0, \SI{5.0}{m/s}] in the random direction.
If the direction of the sampled movement velocity differed by more than 90 degrees from the facing direction, the speed range was reduced to [1.0, \SI{2.5}{m/s}].
These targets were changed 1,004 times, and to account for potential instability during the initial movement phase, measurements began from the fourth goal reset. All comparison models were evaluated using the same test case generated in this manner. The simulation started with the character in a default standing pose.

We defined the following metrics to evaluate the Move policy:
\begin{description}
\item[Moving Goal Achievement Rate (MGAR)]
A goal is considered achieved if, after the goal is set and before the next goal reset, the magnitude of the difference between the character's root horizontal velocity and the target velocity falls within 10\% of the target speed, and the horizontal frontal direction of the character falls within 20 degrees of the target direction.
The metric is defined as the number of achieved goals divided by the number of evaluated goals.

\item[Goal Matching Latent Similarity (GMLS)]
The metric calculates the average cosine similarity.
First, the current input goal of the Move policy is compared to goals in the training data to find the closest match. The motion clip corresponding to this closest goal is passed through the encoder to produce a latent vector.
The cosine similarity between this and the latent vector output by the Move policy is calculated for every frame, and the average is taken.

\end{description}

\begin{figure}
  \centering
  \subfigure[Move-Ours (backward movement)]{
      \includegraphics[trim=0 270 0 330, clip, width=1.\linewidth]{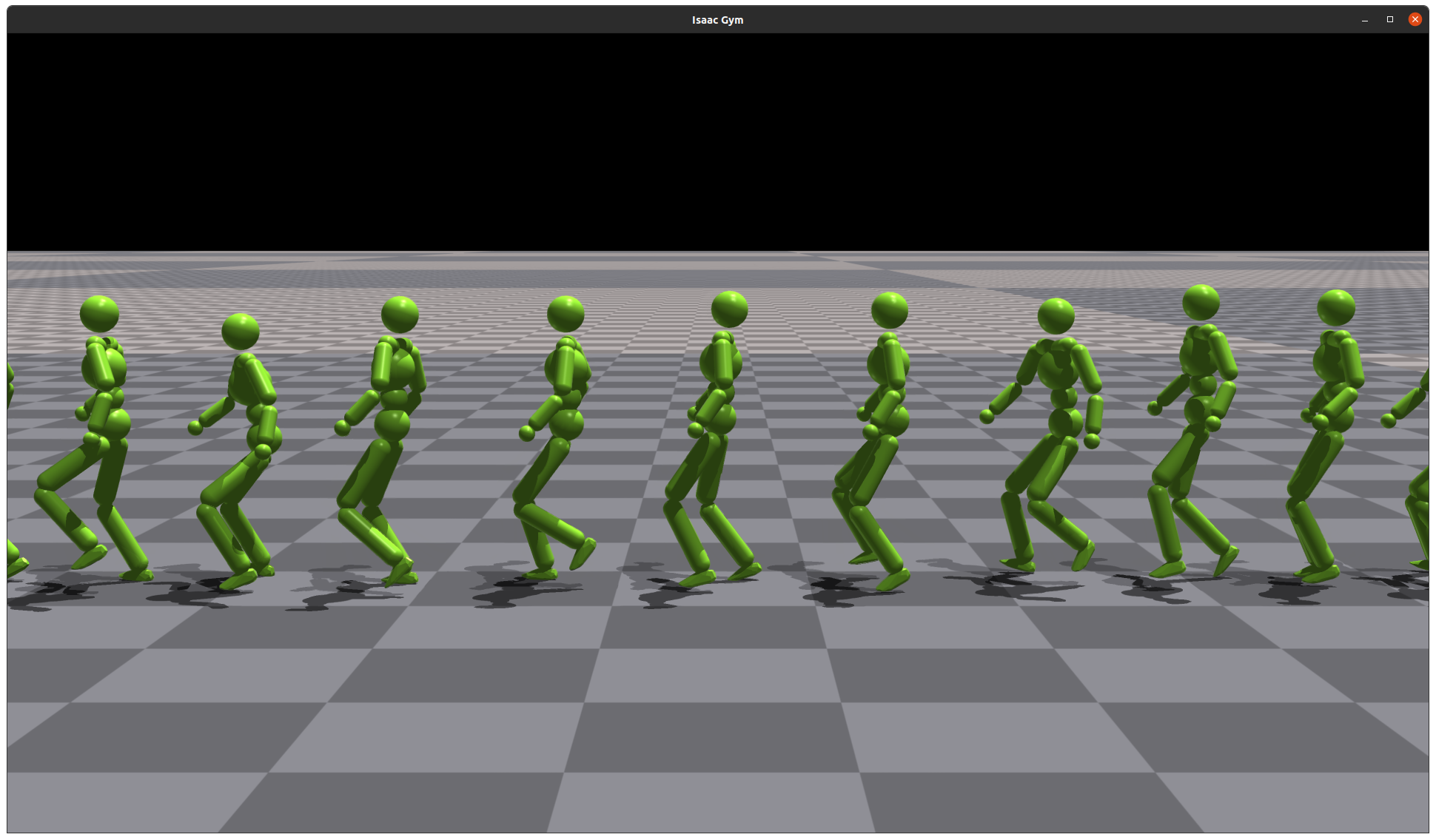}
  }
  \subfigure[Move-w/o DEGCL (backward movement)]{
      \includegraphics[trim=0 270 0 330, clip, width=1.\linewidth]{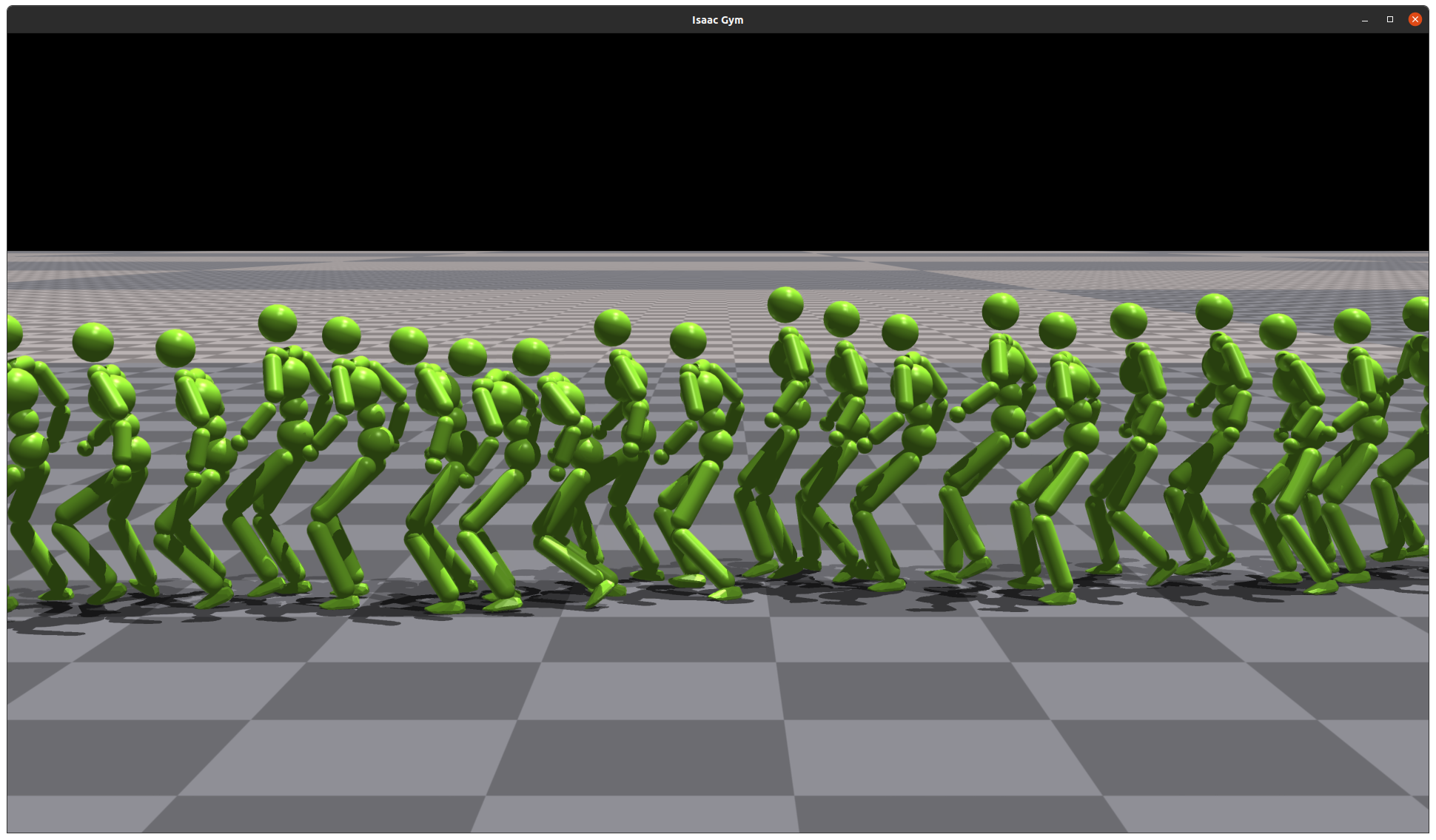}
  }
  \subfigure[Move-Ours (sideways movement)]{
      \includegraphics[trim=0 270 0 330, clip, width=1.\linewidth]{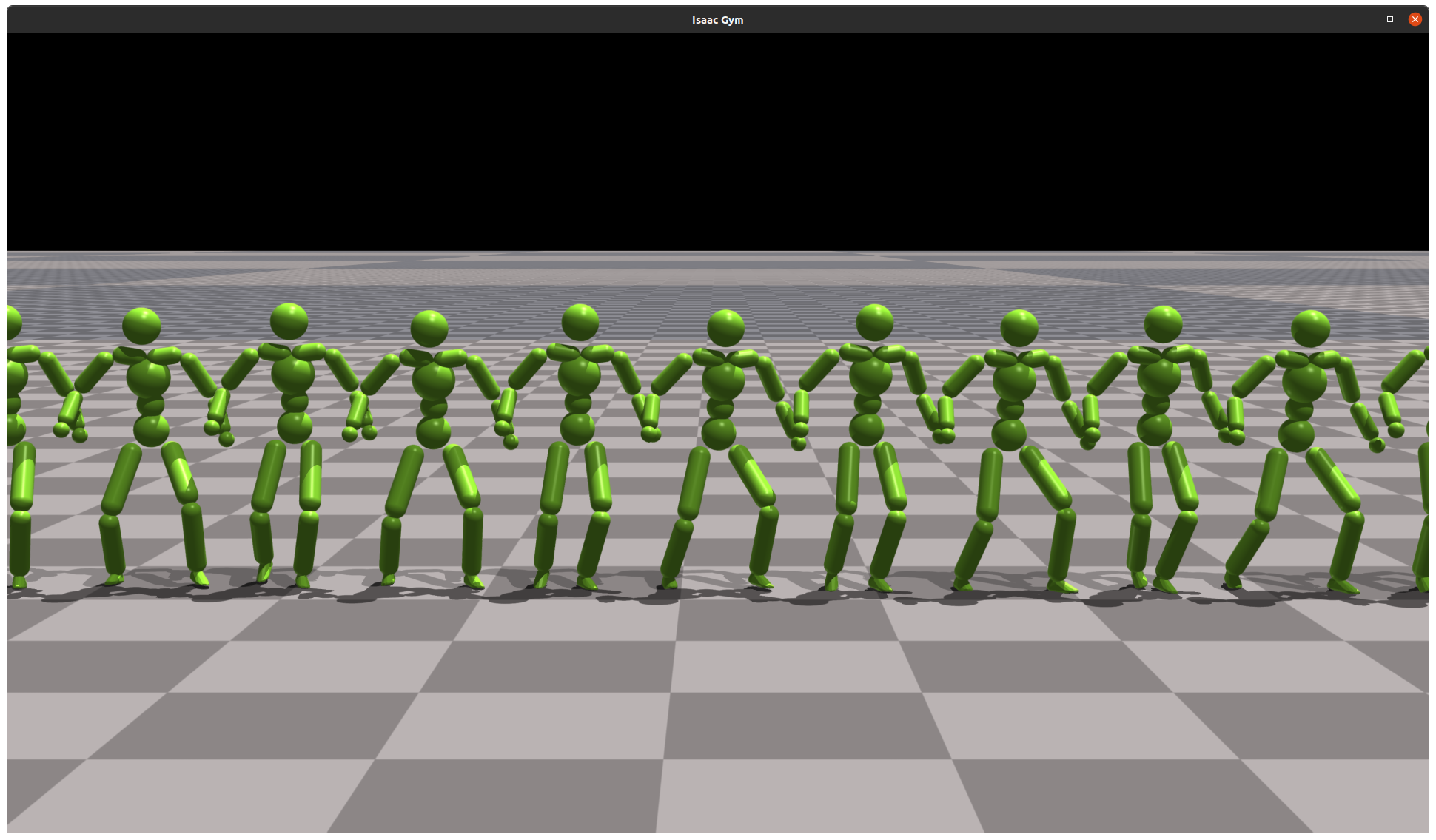}
  }
  \subfigure[Move-w/o DEGCL (sideways movement)]{
      \includegraphics[trim=0 270 0 330, clip, width=1.\linewidth]{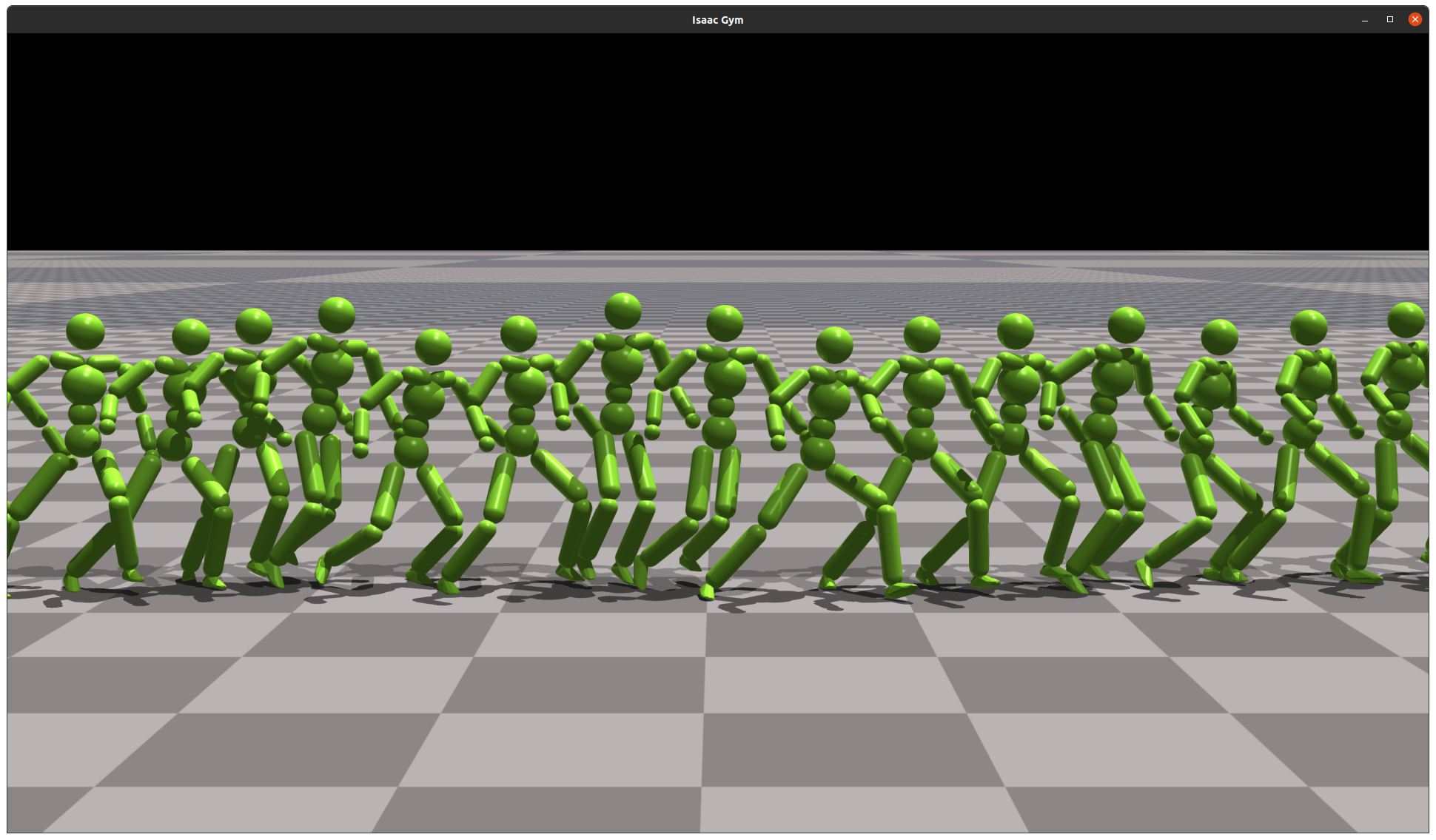}
  }
  \caption{Effect of DEGCL.
(a), (b): For backward movement, the w/o DEGCL model shows excessively short step lengths and high step frequency compared to Ours.
(c), (d):   For sideways movement, the w/o DEGCL model repeatedly exhibits a posture that appears as if the character is about to fall sideways.
In all figures, the character moves from left to right.}
  \label{fig:result-degcl}
\end{figure}

We evaluated the metrics for Ours and the following ablated models:
\begin{description}
\item[Move-w/o $r^{\mathrm{dir}}_t$, $r^{\mathrm{vel}}_t$]
The Move policy trained by excluding each reward term from Equation~\ref{eq:r_move_task}.

\item[Move-w/o DEGCL]
The Move policy trained without using Data-Embedded Goal-Conditioned Latent Guidance, meaning all episodes used for training are general episodes.

\item[Move-w/o NTS]
The Move policy trained without normalization by target speed.

\end{description}

\begin{table}[h]
    \centering
    \caption{Comparison of our Move policy and its ablated models.}
    \begin{tabular}{lccc}
        \toprule
        & MGAR(\%)$\uparrow$ & GMLS$\uparrow$ \\
        \midrule
        Move-w/o DEGCL & \textbf{88.6} & 0.52  \\
        Move-w/o $r^{\mathrm{dir}}_t$ & 41.5 & \textbf{0.63} \\
        Move-w/o $r^{\mathrm{vel}}_t$ & 8.2 & 0.60 \\
        Move-w/o NTS & 82.8 & 0.59  \\
        Move-Ours & 87.9 & 0.62 \\
        \bottomrule
    \end{tabular}
    \label{tbl:move}
\end{table}

As shown in Table~\ref{tbl:move}, models trained with either of the two task reward terms removed exhibit a significant drop in goal achievement rate. The goal achievement rate is slightly highest in the w/o DEGCL model, as this setting excludes the latent similarity reward, allowing the policy to focus solely on the task rewards.
However, as a trade-off, the w/o DEGCL model frequently achieves goals with unnatural motions that deviate from the motions in the training data. This is suggested by Figure~\ref{fig:result-degcl} and the lowest GMLS score recorded by the model. In contrast, Ours and other comparison models with DEGCL applied all achieved GMLS scores around or above 0.6.
It is worth noting that GMLS scores typically range around 0.4 when simulated motions significantly deviate from the specified motions, and around 0.75 when the motions are very close to the specified ones. Thus, even a difference of about 0.1 represents a substantial gap.
The w/o NTS model shows lower performance compared to Ours, although the difference was less pronounced than in the case of the Dribble policy.
Ours recorded a slightly lower MGAR than w/o DEGCL and achieved the highest level of GMLS.

\subsection{Kick}

To evaluate the performance of the Kick policy, we conducted an experiment where kicks were performed according to 1000 randomly assigned target kick velocities.
The target kick velocity was set in the same manner as during the training of the Kick policy: the horizontal direction was uniformly sampled from [-45, $45^\circ$] relative to the character’s forward direction, the vertical direction from [0, $45^\circ$], and the speed from [5, \SI{35}{m/s}].
All comparison models were evaluated using the same test cases consisting of these 1000 target kick velocities. The character’s initial state was set to a default standing pose, and the ball’s initial position was 1 meter in front of the character.

We defined the following metrics to evaluate the Kick policy:

\begin{description}

\item[Kick Success Rate (KSR)]
A kick attempt is considered successful if there is a collision between the character and the ball during the attempt. The metric is defined as the number of kick attempts that touched the ball divided by the total number of kick attempts.

\item[Kick Direction Deviation (KDD)]
The average difference between the actual ball movement direction and the specified target direction during the first 1/6 second after the kick touch. This metric is calculated only for successful kick attempts.

\item[Kick Speed Deviation (KSD)]
The average difference between the actual ball speed and the specified target speed during the first 1/6 second after the kick touch. This metric is calculated only for successful kick attempts.

\end{description}

We measured the metrics for Ours and the following ablated models:
\begin{description}
\item[Kick-w/o NTS]
The Kick policy trained without normalization by target speed.

\item[Kick-w/ BoxFoot]
The Kick policy learned by replacing the character’s foot with a simple box-shaped foot instead of a football boots mesh.

\end{description}

\begin{table}[h]
    \centering
    \caption{Comparison of our Kick policy and its ablated models.}
    \begin{tabular}{lcccc}
        \toprule
        & KSR(\%)$\uparrow$ & KDD($^\circ$)$\downarrow$ & KSD(m/s)$\downarrow$ \\
        \midrule
        Kick-w/o NTS & 0.0 & - & - \\
        Kick-w/ BoxFoot & 99.7 & 6.62 & 6.35 \\
        Kick-Ours & \textbf{99.9} & \textbf{4.79} & \textbf{4.51} \\
        \bottomrule
    \end{tabular}
    \label{tbl:kick}
\end{table}

As shown in Table~\ref{tbl:kick}, NTS played a critical role in the performance of the Kick policy. The w/o NTS model failed to touch the ball even once across 1000 kick attempts, making it impossible to measure the metrics. This result highlights that Kick was significantly more affected by NTS than Dribble or Move.
One possible reason is the broader target speed range for Kick ([5, \SI{35}{m/s}]) compared to Move or Dribble ([0, \SI{7}{m/s}]). This broader range likely made it more challenging to tune the reward coefficient to obtain a valid derivative of the exponential function across the entire range.
The Ours model demonstrated smaller deviations relative to target values compared to the w/ BoxFoot model.

\section{Skill Transition Evaluation}
\label{sec:transition-eval}

To validate the effectiveness of the proposed Skill Transition-Based State Initialization (STI), we quantitatively compared the performance of post-transition skill policies trained with STI (Ours) and without STI (w/o STI) for key transitions where STI is applied, among those defined in the PhysicsFC FSM (Figure~\ref{fig:fsm}).

For each evaluated transition type, we applied the same test cases, consisting of 1000 randomly generated transitions, to both the Ours and w/o STI post-transition policies. The values reported in the tables represent the averages over these 1000 transitions.
In each transition case, the pre-transition policy performed the same assigned target starting from the same initial state and continued until the transition point, ensuring that the simulation before the transition was identical for both Ours and w/o STI.
After the transition point, the simulation behavior began to diverge, and all metrics were measured from this point onward.
All w/o STI post-transition policies were trained with the character's initial state set to a default standing pose.

\subsection{Trap to Dribble}
\label{sec:trap-to-dribble}

At the start of each transition case, the states of the ball and character were initialized in the same manner as the episode initialization used for training the Trap policy, with only the lob pass scenarios.
The pre-transition policy, Trap policy, was provided with the right foot as the target body part.
The post-transition policy, Dribble policy, was given a target dribble velocity as input, consisting of a random direction and a speed randomly sampled from [1, \SI{7}{m/s}].
The Trap-to-Dribble transition occurs when the ball collides with the character, as defined in the PhysicsFC FSM.

To effectively evaluate cases where the post-transition skill is Dribble, we defined and measured the following additional quantitative metrics:

\begin{description}
\item[Dribbling Goal Achievement Rate in 30 seconds (DGAR30)]
A goal is considered achieved if, within 30 seconds after switching to the Dribble policy, the magnitude of the difference between the ball's velocity and the target velocity falls within 10\% of the target speed.

\item[Time to Achieve Dribbling Goal (TADG)]
The time taken to achieve the goal after transitioning to the Dribble policy. This is measured only for cases where the goal is achieved within 30 seconds, based on the DGAR30 criterion.

\end{description}

\begin{table}[h]
    \centering
    \caption{Comparison of Trap-to-Dribble transition performance with and without STI.}
    \begin{tabular}{lccc}
        \toprule
         & TADG(s)$\downarrow$ & DGAR30(\%)$\uparrow$ \\
        \midrule
        Dribble-Ours & \textbf{2.71} & \textbf{94.0} \\
        Dribble-w/o STI & 3.54 & 92.2 \\
        \bottomrule
    \end{tabular}
    \label{tbl:trap-to-dribble}
\end{table}

As shown in Table~\ref{tbl:trap-to-dribble}, the Dribble policy trained with STI achieved dribbling goals approximately 77\% faster than the policy trained without STI. Additionally, it recorded a slightly higher goal achievement rate.
This indicates that STI plays a crucial role in enabling the Dribble policy to quickly control the ball as desired. The differences in the transition behaviors generated by the two models can be observed in detail in the accompanying video.

\subsection{Move to Dribble}
\label{sec:move-to-dribble}

At the start of each transition case, the character's state was initialized to a default standing pose.
The ball's initial position was randomized within a radius of [2.5, \SI{3.5}{m}] around the character, and its initial velocity was randomly set to a speed between [0, \SI{1}{m/s}] in a random direction.
For the pre-transition policy, Move policy, the goal input at each timestep consisted of a target facing direction and a target movement velocity direction, both pointing toward the current position of the ball. The magnitude of the target movement velocity was randomly selected from the range [1, \SI{7}{m/s}].
The Move-to-Dribble transition occurred when the horizontal distance between the ball and the character’s root became less than \SI{2}{m}, as defined in the PhysicsFC FSM.

\begin{table}[h]
    \centering
    \caption{Comparison of Move-to-Dribble transition performance with and without STI.}
    \begin{tabular}{lccc}
        \toprule
         & TADG(s)$\downarrow$ & DGAR30(\%)$\uparrow$ \\
        \midrule
        Dribble-Ours & \textbf{1.51} & \textbf{88.7} \\
        Dribble-w/o STI & 2.80 & 88.6 \\
        \bottomrule
    \end{tabular}
    \label{tbl:move-to-dribble}
\end{table}

As shown in Table~\ref{tbl:move-to-dribble}, the Dribble policy trained with STI recorded a goal achievement rate similar to that of the w/o STI policy but achieved dribbling goals approximately 54\% faster.
Similar to the Trap-to-Dribble transition, this indicates that STI plays a crucial role in enabling a swift transition to dribbling.

\subsection{Move to Trap}

At the start of each transition case, the character's state was initialized in the same manner as in the Move-to-Dribble evaluation (Section~\ref{sec:move-to-dribble}), while the ball's state was initialized in the same way as in the Trap-to-Dribble evaluation (Section~\ref{sec:trap-to-dribble}).
The ball was launched after a randomly selected time in the range of [1, \SI{2}{s}] following the execution of the Move policy.
The goal input for the pre-transition policy, Move policy, was also provided in the same manner as in the Move-to-Dribble evaluation. The Move-to-Trap transition occurred at the moment the ball was launched.

\begin{table}[h]
    \centering
    \caption{Comparison of Move-to-Trap transition performance with and without STI.}
    \begin{tabular}{lccc}
        \toprule
         & TSR(\%)$\uparrow$ & RBSPT(m/s)$\downarrow$ \\
        \midrule
        Trap-Ours & \textbf{74.1} & \textbf{4.85} \\
        Trap-w/o STI & 55.1 & 5.12 \\ 
        \bottomrule
    \end{tabular}
    \label{tbl:move-to-trap}
\end{table}

As shown in Table~\ref{tbl:move-to-trap}, the Trap policy trained with STI achieved a TSR approximately 19\% higher than the w/o STI policy. Additionally, the ball's speed immediately after the touch was measured to be lower, indicating that STI plays a crucial role in successfully performing trapping during movement.

\subsection{Dribble to Kick}

At the start of each transition case, the states of the character were initialized in the same manner as the episode initialization used during the training of the Dribble policy.
The initial position of the ball is \SI{1.5}{m} in front of the character.
For the pre-transition policy, Dribble policy, the goal input was a target dribble velocity with a magnitude randomly sampled from the range [1, \SI{7}{m/s}] in the character's forward direction.
For the post-transition policy, Kick policy, the target kick velocity was randomly sampled within a direction range of [-45, $45^\circ$] relative to the character's forward direction and a speed range of [7, \SI{30}{m/s}].
The Dribble-to-Kick transition occurred after a randomly selected time within the range of [3, \SI{5}{s}] following the execution of the Dribble policy.

To effectively evaluate this type of skill transition, we defined and measured the following additional quantitative metrics:
\begin{description}

\item[Time to Kick (TTK)]
The time taken from transitioning to the Kick policy until the ball collides with the character's foot.

\end{description}

\begin{table}[h]
    \centering
    \caption{Comparison of Dribble-to-Kick transition performance with and without STI.}
    \begin{tabular}{lcccc}
        \toprule
         & KSR(\%)$\uparrow$ & TTK(s)$\downarrow$ &  KDD($^\circ$)$\downarrow$ & KSD(m/s)$\downarrow$ \\
        \midrule
        Kick-Ours & \textbf{100} & \textbf{2.99} & \textbf{16.9} & \textbf{5.81} \\
        Kick-w/o STI & 16.95 & 3.35 &  37.41 & 7.21 \\
        \bottomrule
    \end{tabular}
    \label{tbl:dribble-to-kick}
\end{table}

As shown in Table~\ref{tbl:dribble-to-kick}, the use of STI was found to have a critical impact on the Dribble-to-Kick transition. The Kick-Ours policy, trained with STI, successfully kicked the ball in all 1,000 attempts without a single failure. In contrast, the Kick-w/o STI policy, which never encountered situations involving kicking during dribbling during training, was able to kick the ball in only about 17\% of the 1,000 Dribble-to-Kick attempts.
Even in the successful 17\% of cases, the time taken to actually perform the kick after transitioning to the Kick policy was approximately 12\% longer compared to the STI-trained policy. Additionally, the difference between the actual ball velocity and the target kick velocity was measured to be 1.2 times greater (KSD) to as much as 2.2 times greater (KDD) than Ours.

\section{Discussion}

In this paper, we introduced PhysicsFC, a method for controlling physically simulated football player characters to perform a range of football skills—including dribbling, trapping, kicking, and moving—based on user input, while seamlessly transitioning between these skills. By leveraging a hierarchical framework with skill-specific policies trained on a physics-based motion embedding model, PhysicsFC enables realistic, agile, and context-appropriate football movements in a simulated environment. The proposed system incorporates innovative techniques, such as tailored reward designs for skill training, Data-Embedded Goal-Conditioned Latent Guidance (DEGCL) for movement diversity, and Skill Transition-Based Initialization (STI) for smooth transitions. Through interactive demonstrations and quantitative evaluations, we demonstrated the system's potential for generating user-controllable, physics-based football gameplay that bridges the gap between realism and interactivity.

While the proposed PhysicsFC represents a meaningful step forward in this direction, several limitations remain to be addressed.
One notable limitation is that the Dribble policy consistently learns to rely almost exclusively on a single foot (e.g., the left foot) for ball touches during dribbling, and the same applies to the Kick policy.
This behavior likely stems from the character initially experiencing changes in the ball's velocity by touching it with one foot, which leads to further learning of controlling the ball using the same foot throughout training.
Addressing this limitation by encouraging the character to use both feet for ball control, or by allowing the dribbling or kicking foot to be specified, could enhance the realism and versatility of PhysicsFC.
Another limitation is that the Magnus effect, which affects the curvature of a ball’s trajectory during flight, is not accounted for in the physics simulation.
This omission limits the ability to accurately reflect the realistic movement of the ball, particularly for high-speed or spinning balls.
Incorporating the Magnus effect into the training of Kick or Trap policies could be a valuable future direction that can enable more realistic simulated football gameplay.
\textcolor{rv}{Fall recovery was trained similarly to ASE~\cite{ASE} and CALM~\cite{CALM}, starting from random fallen states. This approach often led to standing motions with high joint torques and abrupt transitions. Incorporating motion data or human-inspired models, such as musculoskeletal systems 
\cite{lee2014locomotion,feng_musclevae_2023}, 
could further improve realism.}

Future work could explore several avenues to enhance the realism and versatility of the proposed system. Introducing a broader range of dribbling styles, such as inside-foot and outside-foot touches, as well as advanced techniques like feints and step-overs, would allow the simulated characters to exhibit a more diverse and lifelike repertoire of movements. Additionally, incorporating competitive scenarios, including interactions with defenders or contested ball situations, could enable the system to handle more dynamic and realistic football contexts. Expanding the range of football skills and testing the system in these challenging scenarios would further bridge the gap between simulation and real-world football dynamics.

\begin{acks}
This work was supported by the National Research Foundation of
Korea (NRF) grant (RS-2023-00222776); by Culture, Sports and Tourism R\&D Program through the Korea Creative Content Agency grant funded by the Ministry of Culture, Sports and Tourism in 2024 (RS-2024-00399136); and by Institute of Information \& communications Technology Planning \& Evaluation (IITP) grant funded by the Korea government (MSIT) (RS-2020-II201373, Artificial Intelligence Graduate School Program(Hanyang University)).
\end{acks}

\bibliographystyle{ACM-Reference-Format}
\bibliography{main}

\appendix

\section{Physics-Based Motion Embedding Model: CALM}
\label{sec:appd-calm}

Recently, latent representation-based approaches have gained significant attention as a method for enabling physically simulated characters to perform diverse motions and their variations from motion datasets ~\cite{ASE,won_physics-based_2022,yao_controlvae_2022,dou_case_2023,CALM,zhu_neural_2023,yao_moconvq_2024}.
These approaches focus on learning a shared latent representation that can be applied across various downstream tasks.
Specifically, they utilize an encoder to embed the diverse motions in a motion dataset into a low-dimensional latent space, paired with a low-level control policy that takes a latent variable $\mathbf z$ sampled from this space and physically reproduces the corresponding motion.
This framework allows for the training of a high-level control policy that outputs latent variables $\mathbf z$ to perform various actions required for specific downstream tasks.
Since this approach maps each motion segment in the dataset into a corresponding latent $\mathbf z$, which allows the low-level policy to physically reproduce the motion, we refer to it in this paper as a \textit{physics-based motion embedding model}.

\begin{figure}[h]
  \centering
  \includegraphics[trim=130 70 150 70, clip, width=1.\linewidth]{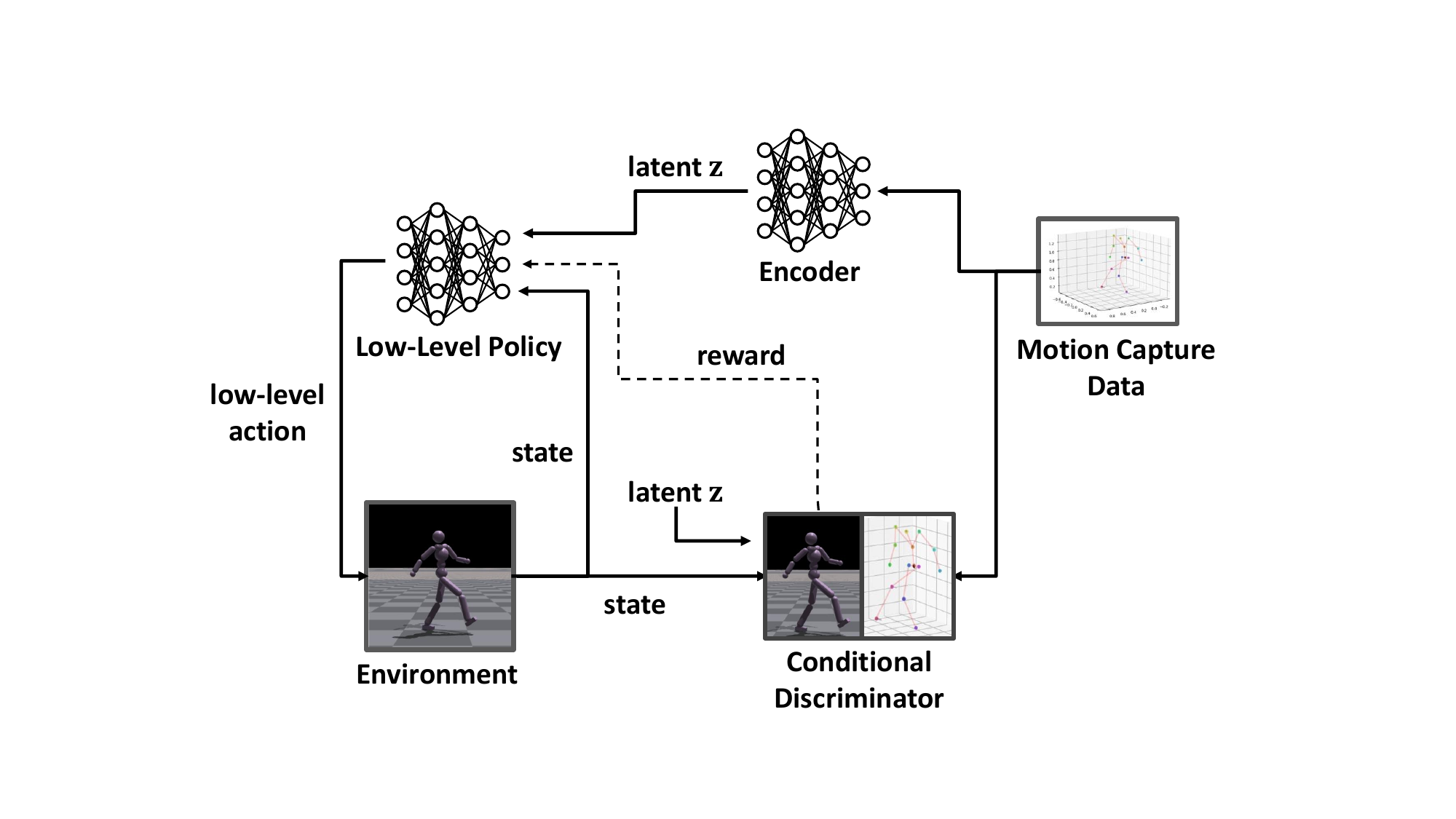}
  \caption{Structure of CALM model}
  \label{fig:calm}
\end{figure}

We utilize CALM~\cite{CALM} as a physics-based motion embedding model, which trains the low-level policy, encoder, and conditional discriminator (Figure~\ref{fig:calm}).
The encoder converts high-dimensional motion sequences into low-dimensional latent variables $\mathbf z$, while the low-level policy takes $\mathbf z$ from the encoder, along with the character's state, and generates the low-level actions needed for simulation.
The discriminator, conditioned on $\mathbf z$, is trained to differentiate between simulated motion and the original motion sequences.
At the same time, the low-level policy is rewarded for producing low-level actions that generate simulated motion closely resembling the original motion, effectively "fooling" the discriminator.
This reward signal enables the encoder and low-level policy to be trained end-to-end.
The character state input to the low-level policy is in the same format as the character state used in the skill policies, as described in Appendix~\ref{sec:appd-character-ball-states}.
The latent variable $\mathbf z$ is 64-dimensional.

\section{Character and Ball States}\label{sec:appd-character-ball-states}

The character state and ball state, which are common inputs to all our skill policies, are structured as follows:

Character state $\in \mathbb R^{223}$: 
\begin{itemize}
    \item Root(pelvis) height $\in \mathbb R^{1}$
    \item Body part positions (excluding root) $\in \mathbb R^{14 \times 3}$
    \item Body part rotations $\in \mathbb R^{15 \times 6}$
    \item Body part linear velocities $\in \mathbb R^{15 \times 3}$
    \item Body part angular velocities $\in \mathbb R^{15 \times 3}$
\end{itemize}

Ball state $\in \mathbb R^{13}$:
\begin{itemize}
    \item Ball position $\in \mathbb R^{3}$
    \item Ball rotation $\in \mathbb R^{4}$
    \item Ball linear velocity $\in \mathbb R^{3}$
    \item Ball angular velocity $\in \mathbb R^{3}$
\end{itemize}

All the above elements are expressed in the current character coordinate system, where the character root's forward direction defines the x-axis, the global upward vertical direction defines the z-axis, their cross product defines the y-axis, and the root position projected onto the ground serves as the origin.

\section{Dribble Reward Details}
\label{sec:appd-dribble-reward}

The dribbling policy is trained using the following rewards:
\begin{equation}
\begin{aligned}
r^{\mathrm{drib}}_t = 
0.6 \ r^{\mathrm{ball\_vel}}_t +
0.2 \ r^{\mathrm{ball\_root\_pos}}_t + 
0.2 \ r^{\mathrm{root\_vel}}_t.
\label{eq:r_drib_apdx}
\end{aligned}
\end{equation}

$r^{\mathrm{ball\_vel}}_t$ encourages the ball to move at the target dribble velocity:
{\small
\begin{equation}
r^{\mathrm{ball\_vel}}_t = \
\mathrm{exp} \left(-10 \left( \left( \frac {\| \hat{\mathbf v}^{\mathrm{drib}}_t-\mathbf v^{\mathrm{ball(2)}}_t \|} {\|\hat{\mathbf v}^{\mathrm{drib}}_t\| +  \epsilon} \right)^2 + 0.1 \left( \frac {\| \hat{\mathbf v}^{\mathrm{drib}}_t \| - \| \mathbf v^{\mathrm{ball(2)}}_t \|} {\|\hat{\mathbf v}^{\mathrm{drib}}_t\| +  \epsilon} 
\right)^2 \right) \right),
\label{eq:r_ball_vel}
\end{equation}
}
where $\hat{\mathbf v}^{\mathrm{drib}}_t$ represents the target dribble velocity and $\mathbf v^{\mathrm{ball(2)}}_t$ denotes the horizontal velocity of the ball.
This reward term enables the Dribble policy to learn to dribble the ball in the target direction and at the target speed.

\paragraph{
\textbf{Normalization by Target Speed (NTS)}}
Both terms inside the exponential function in Equation~\ref{eq:r_ball_vel} are normalized by the target speed (NTS).
This normalization ensures consistent reward signals regardless of the range or magnitude of the target speed, reducing the need for extensive coefficient tuning. 

Without NTS, when the difference between the current and target speeds is large, the derivative of the exponential function becomes very small.
In such cases, small improvements in the gap between the current and target speeds have minimal impact on the reward, making it challenging for the policy to learn effectively.
By normalizing with the target speed, NTS ensures that the exponential function remains sensitive to such large differences, which are commonly observed during the early stages of learning at higher target speeds. 
This means that small reductions in the gap between the current and target speeds result in a more noticeable reward increase, encouraging the policy to continue improving.
Importantly, this reduces the need for extensive tuning of the coefficients inside the exponential function, allowing the policy to learn consistently across a wide range of target speeds.

\textcolor{rv}{CompositeMotion~\cite{xu_composite_2023} applies standardization to each objective term based on sample variance during training. 
In contrast, our NTS method focuses on normalizing the reward specifically by the target speed, addressing the unique challenges of dynamic football gameplay where target speeds vary widely.}

In addition, in Equation~\ref{eq:r_ball_vel}, the difference in speed is calculated separately from the difference in velocity to encourage the policy to better satisfy the target speed.
To prevent division by zero, a small constant $\epsilon=0.01 \text{m/s}$ was added.

$r^{\mathrm{ball\_root\_pos}}_t$ encourages the character's root (pelvis) and the ball to maintain a close distance on the horizontal plane:
\begin{equation}
r^{\mathrm{ball\_root\_pos}}_t = \
\mathrm{exp} \left(-10 \left\| \mathbf x_t^{\mathrm{ball(2)}}-\mathbf x_t^{\mathrm{root(2)}} \right\|^2 \right), \\
\end{equation}
where $\mathbf x^{\mathrm{root(2)}}_t$ and $\mathbf x^{\mathrm{ball(2)}}_t$ represent the horizontal positions of the character's root (pelvis) and the ball, respectively.
During dribbling, the character's feet repeatedly move closer to and farther from the ball.
Therefore, we use the character's root instead of its feet.
If the horizontal distance between the character's root and the ball becomes too short, causing the ball to be positioned between the character's feet, proper dribbling becomes impossible, leading to lower values for the other reward terms in Equation~\ref{eq:r_drib_apdx}.
Thus, the policy is trained to keep the ball in front of the character while moving, resulting in the character dribbling while keeping the ball close to its feet.

$r^{\mathrm{root\_vel}}_t$ encourages the character to move toward the current position of the ball at the target dribble speed:
{\small
\begin{equation}
r^{\mathrm{root\_vel}}_t \!=\! 
\mathrm{exp} \left(-10 \left( \left( \frac {\left\|  \|\hat{\mathbf v}^{\mathrm{drib}}_t\| \mathbf d^{\mathrm{r2b}}_t-\mathbf v^{\mathrm{root(2)}}_t \right\|} {\|\hat{\mathbf v}^{\mathrm{drib}}_t\| +  \epsilon } \right)^2 \!+\! 0.1 \left( \frac {\| \hat{\mathbf v}^{\mathrm{drib}}_t \| - \| \mathbf v^{\mathrm{root(2)}}_t \|} {\|\hat{\mathbf v}^{\mathrm{drib}}_t\| +  \epsilon} \right)^2 \right) \right), \\
\end{equation}
\begin{equation}
\mathbf d^{\mathrm{r2b}}_t=\frac{\mathbf x_t^{\mathrm{ball(2)}}-\mathbf x_t^{\mathrm{root(2)}}}{\| \mathbf x_t^{\mathrm{ball(2)}}-\mathbf x_t^{\mathrm{root(2)}} \|},
\end{equation}
}
where $\mathbf v^{\mathrm{root(2)}}_t$ represents the horizontal velocity of the character's root, and $\mathbf d^{\mathrm{r2b}}_t$ is the unit direction vector from the character's root to the ball on the horizontal plane.
This reward term, also normalized by the target speed, encourages the character to consistently move toward the ball, enabling the policy to learn to dribble without losing control of the ball.

\section{Relationship Between Ball's Launch Angle, Initial Distance, and Flight Time}\label{sec:appd-ball-distance-angle}

When the height of the ball's initial position $\mathbf p_\text 0$ equals the height of the landing position $\mathbf p_\text f$ ($\mathbf p_\text{f,z} = \mathbf{p}_\text{0,z}$), we can derive the distance $d$ between $\mathbf p_\text 0$ and $\mathbf p_\text f$,  vertical launch angle $\phi$, and the flight time $t$ required for the ball, launched with an initial speed $v_0$, to land precisely at $\mathbf p_\text f$.
This derivation assumes that the ball's trajectory is influenced solely by gravity, with no external factors such as wind or the Magnus effect impacting its motion.

The equation describing the ball's vertical motion is as follows:
\begin{equation}
\mathbf p_\text{f,z} = \mathbf{p}_\text{0,z} + v_0 \sin \phi \cdot t - \frac{1}{2} g t^2,
\label{eq:ball-vertical}
\end{equation}
where $t$ represents the time elapsed since the ball was launched, and $g$ denotes the gravitational acceleration (\SI{9.8}{m/s^2}).
Given the condition $\mathbf p_\text{f,z} = \mathbf{p}_\text{0,z}$, Equation~\ref{eq:ball-vertical} simplifies as follows:
\begin{equation}
0 = v_0 \sin \phi \cdot t - \frac{1}{2} g t^2,
\end{equation}

The solution for $t \neq 0$ is as follows:
\begin{equation}
t = \frac{2 v_0 \sin \phi}{g}.
\label{eq:ball-t}
\end{equation}

The distance $d$ between $\mathbf p_\text 0$ and $\mathbf p_\text f$ is the horizontal distance traveled at the initial horizontal velocity $v_0 \cos \phi$ over time $t$:
\begin{equation}
d = v_0 \cos \phi \cdot t = v_0 \cos \phi \cdot \frac{2 v_0 \sin \phi}{g} = \frac{2 v_0^2 \sin \phi \cos \phi}{g}.
\label{eq:ball-d}
\end{equation}

By applying the sine double angle identity $\sin 2\phi = 2 \sin \phi \cos \phi$, Equation~\ref{eq:ball-d} can be simplified as follows:
\begin{equation}
v_0^2 = \frac{d g}{\sin 2\phi}
\label{eq:ball-distance-angle}
\end{equation}

Equation~\ref{eq:ball-distance-angle} describes the relationship between the distance $d$ and the vertical launch angle $\phi$ for a given initial speed $v_0$, while the corresponding flight time $t$ can be determined using Equation~\ref{eq:ball-t}.

\section{Move Task Reward Details}
\label{sec:appd-move-task-reward}

The task reward for the move policy is defined as follows:
\begin{equation}
r^{\mathrm{mv\_task}}_t = 
0.7 \ r^{\mathrm{vel}}_t + 
0.3 \ r^{\mathrm{dir}}_t.
\end{equation}

$r^{\mathrm{vel}}_t$ encourages the character root to move at the target movement velocity:
{\small
\begin{equation}
r^{\mathrm{vel}}_t = \
\mathrm{exp} \left(-0.25 \left( \left( \frac {\| \mathbf v^{\mathrm{target}}_t - \mathbf v^{\mathrm{root(2)}}_t \|} {\| \mathbf v^{\mathrm{target}}_t \| +  \epsilon} \right)^2 + 0.1 \left( \frac {\| \mathbf v^{\mathrm{target}}_t \| - \| \mathbf v^{\mathrm{root(2)}}_t \|} {\| \mathbf v^{\mathrm{target}}_t \| +  \epsilon} 
\right)^2 \right) \right),
\label{eq:r_move_vel}
\end{equation}
}
where $\mathbf v^{\mathrm{target}}_t$ denotes the target movement velocity, and $\mathbf v^{\mathrm{root(2)}}_t$ represents the horizontal velocity of the character root.
This equation, like Equation~\ref{eq:r_ball_vel}, includes a speed difference term and is normalized by the target speed.

$r^{\mathrm{dir}}_t$ encourages the facing direction of the character root to align with the target facing direction:
\begin{equation}
r^{\mathrm{dir}}_t = \mathbf d^{\mathrm{target}}_t \cdot \mathbf d^{\mathrm{root}}_t,
\end{equation}
where $\mathbf d^{\mathrm{target}}_t$ denotes the target facing direction unit vector, and $\mathbf d^{\mathrm{root}}_t$ represents the facing direction of the character root on the horizontal plane.

\section{Motion Dataset Details}
\label{sec:appd-dataset}

The football motion dataset\footnote{https://assetstore.unity.com/packages/3d/animations/aa-soccer-mega-animations-pack-241419, accessed January 21, 2025} purchased from the Unity Asset Store was utilized for training the low-level policy. The dataset consists of 90 clips, covering most movements seen in football games, such as locomotion, jumping, dribbling, kicking, and passing.

The original motion data, composed of 65 body parts, was retargeted twice for training the low-level policy. First, the original motion was retargeted to the skeleton format of the CMU Motion Capture Database, which consists of 38 body parts, using the Rokoko Studio Live Plugin for Blender. Subsequently, it was further retargeted to the skeleton used in CALM \cite{CALM}, which consists of 15 body parts, utilizing the poselib library included in CALM's public implementation.

The final motion data used for training comprised a total of 11,318 frames, equivalent to 188 seconds of motion. The clips ranged in length from 0.3 seconds to 2.5 seconds and were used as-is, without additional cropping, for training the low-level policy.

For training the high-level policies for each football skill, the already-trained low-level policy was employed, and the motion data was not used directly.

\section{Network Architecture and Training Time}
\label{sec:appd-network}

The low-level policy, discriminator, and encoder each consist of three fully connected layers with sizes [1024, 1024, 512].
The latent vector $\mathbf z$ has a dimensionality of 64.
Each high-level football skill policy is implemented as a fully connected network with layers of sizes [1024, 512].

\textcolor{rv}{
The 30-day training time of the low-level policy is not an absolute requirement—most motions can be learned within 2–3 weeks. However, we extended training to further stabilize motion reproduction, considering discriminator loss convergence, to enhance motion fidelity and quality. 
}

\section{Physics Simulation Configuration}
\label{sec:appd-simulation}

The metrics we proposed for quantitative evaluation (Section~\ref{sec:skill-eval}, \ref{sec:transition-eval}), such as Foot-Ball Distance (FBD), may yield different measurements depending on the physical simulation settings used (e.g., the ball's friction coefficient). Therefore, to facilitate the comparison of quantitative measurements in future studies, we document the exact physical simulation settings used during the training and evaluation of our policies.

The physics simulation engine we used is Isaac Gym Preview 4 Release. The ground was configured with a friction coefficient of 1.0 and a restitution coefficient of 0.2.
The character was adopted from the public implementation of CALM~\cite{CALM}, with a friction coefficient of 1.0 and a restitution coefficient of 0.0.
The ball was configured with a friction coefficient of 0.2 (including a rolling friction coefficient of 0.2) and a restitution coefficient of 0.8. To account for a very simple air resistance effect, the linear damping coefficient was set to 0.1, and the angular damping coefficient was set to 0.05.
Although this is not officially documented in the Isaac Gym documentation, it has been mentioned in the developer forums\footnote{https://forums.developer.nvidia.com/t/how-to-randomize-ground-plane-friction/187389, accessed January 21, 2025} that the friction coefficient between two objects in Isaac Gym is calculated as the average of their respective friction coefficients. 
Based on this understanding, we believe that the friction coefficients applied to the ball-character, ball-ground, and character-ground interactions are reasonable compared to those observed in real-world scenarios.
The simulation settings configured for each entity are summarized in Table~\ref{tbl:simulation}.

\begin{table}[h]
    \centering
    \caption{Isaac Gym Preview 4 Release simulation settings used for training all policies and runtime simulation.}
    \begin{tabular}{lccc}
        \toprule
        Coefficient & Ball & Ground & Character \\
        \midrule
        Friction &  0.2 & 1.0 & 1.0 \\
        Rolling friction & 0.2 & - & 0.5 \\
        Restitution & 0.8 & 0.2 & 0.0 \\
        Linear damping & 0.1 & - & 0.0 \\
        Angular damping & 0.05 & - & 0.5 \\
        \bottomrule
    \end{tabular}
    \label{tbl:simulation}
\end{table}

\section{Interactive Demo Details}
\label{sec:appd-interactive-demo}

\paragraph{User-Controlled Give and Go Play}
The user-controlled player passes the ball to the other player on the team and runs forward to penetrate the defense of the opposing players. The team player passes back to the user-controlled player after trapping, then the user-controlled player takes the ball and tries to make a goal by kicking the ball.

In the case of controlled player, the goal input of each policy is determined by the user's gamepad input, and switching between policies is done by the PhysicsFC FSM.
The desired kick velocity of the controlled player's first pass toward teammate, which is the input to the Kick policy, is calculated as follows:
For a lob pass, the current position of the ball is set as the launch point, and the current position of the teammate is set as the landing point. Based on user input, the vertical launch angle is determined within a range of \SI{0.45} to \SI{45} degrees, and the initial velocity vector is analytically calculated to achieve the desired parabolic trajectory.
The initial velocity for a ground pass is calculated by adjusting only the vertical component of the target kick velocity vector from the lob pass, so that the vertical launch angle becomes approximately 3 degrees.
In cases other than passes, the input for the kick policy is entirely determined by the user's gamepad input.
When receiving a pass from a teammate, the ball is received either through trapping, based on the user's Trap button input, or through a Move-to-Dribble skill transition.
For both the user-controlled player and team player, the target body part used in the Trap policy is randomly set to one of the two feet.

Basically, a teammate is given an input for the Move policy to run towards the goal line at \SI{3.5}{m/s}, and switches to the Trap policy at the moment the user-controlled player kicks for the first pass.
In the dribbling state, the target position of the team player is set along the vertical line drawn from his current position to the goal line. This position is determined so that their distance to the goal line matches the distance between the user-controlled player and the goal line.
The target dribble velocity is set in the direction toward the goal line, with its magnitude clamped between \SI{2}{m/s} and \SI{3}{m/s}, proportional to the distance difference in the forward direction from the teammate's position to the target position.
The timing passing the ball back to the user-controlled player is determined by the user pressing a specific button. At this moment, the desired kick velocity input for the Kick policy is calculated as a ground pass, with the target position set to a point \SI{1}{m} ahead of the user-controlled player.

For an opposing team player running to intercept the ball, the Move policy input instructs them to run straight toward a position offset from the ball's current position by a distance proportional to the ball's speed in the direction of its movement. Meanwhile, the other opposing player will receive a Move policy input to run toward 8 meters ahead of the target, whichever is closer to the goal line the ball or the user's team player.

\paragraph{Competitive Trapping and Dribbling}

This demo showcases a scenario where the controlled player and an opposing player repeatedly compete for control of the ball by attempting trapping and dribbling.

In the demo, the players transition their skills based on the PhysicsFC FSM, starting in the Move state. At regular intervals, when a new ball is launched, a Trap start command is issued to all players, triggering the corresponding behavior.
The target body part entered into the trap policy is randomly set to one of the five body parts excluding the head in the case of a lob pass, and randomly set to one of the two feet in the case of a ground pass.

Regarding the opposing player, the target movement velocity is set to run at maximum speed (\SI{7}{m/s}) toward the ball if they are within \SI{10}{m}; otherwise, they run at maximum speed toward the user-controlled player. The target facing direction is always set to face the ball.
When dribbling, target dribble velocity is given to run toward the user-controller player's goal at maximum speed. Once the distance to the goal becomes within \SI{15}{m}, a kick start command is dispatched, switching to the Kick policy and providing input to kick toward the goal.

The user-controlled player’s target input for each policy is determined by the user's gamepad input, and transitions between policies are managed by the PhysicsFC FSM.

\paragraph{Simulated 11v11 Football Game with User-Controlled Player Switching}

A total of 22 PhysicsFC agents, consisting of 11 from user team and 11 from the opposing team, are simultaneously simulated in the demo.

The user can designate the target team player for the current user-controlled player to pass to using the right stick on the gamepad. At the moment the kick occurs, control is immediately switched to the target player, who is then provided a Trap start command.
The inputs for the Trap and Kick policies are calculated in the same manner as in the Give and Go Play demo.
If there are no players in the Dribble or Trap state on the user's team, the user-controlled player can be changed using the directional pad on the controller. Additionally, pressing the Y button assigns the team player closest to the ball as the controlled player.
Even without any control switching input, if one or more players on the user's team are in the Dribble or Trap state, control target automatically switches to the player closest to the ball among them.

The players on both the user's team and the opposing team are positioned on the field in a 4-3-1-2 formation.
Each player has their own Move target position based on their assigned position in the formation.
This target position adjusts dynamically, expanding or contracting toward the user's team or the opposing team depending on the ball's position.
For example, if the ball is on the user's team's side of the field, the target positions of all players on both teams move closer to the user's goal line (and vice versa). Additionally, the farther the ball is from the halfway line, the greater the extent of this movement.

Opposing team players are given a Move policy input to move towards the ball if the ball is within \SI{10}{m} of their respective current positions and within \SI{15}{m} of their respective target positions.
If there is no opposing player who satisfies both conditions, a Move policy input is given to cause the closest opposing player to move towards the ball.
The Move policy input for opposing players moving toward the ball is calculated in the same way as the opposing player chasing the ball in the Give And Go Play demo.
Only the opposing player moving toward the ball has their target facing direction set toward the ball, while other opposing players have their target facing direction aligned with their movement direction to quickly reach their target positions.
Players on the user team always have their target facing towards the ball. The goalkeepers use inputs set in the same way as each team's players, except that their formation standard is in front of the goal.

\section{Runtime Gamepad Input}
\label{sec:appd-gamepad}

\begin{table}[H]
    \caption{Gamepad input mapping used in the interactive demos.}
    \begin{tabular}{llll}
        \toprule
        Input & State & Function & Range \\
        \midrule
        Left trigger & Dribble  & Kick start command      & -\\
                     & Kick     & Lob pass vertical angle &$\left(0.45, 45\right]\mathbf{^\circ}$ \\
        \midrule
        Right trigger& Kick     & Target kick speed       & $\left[5, 35\right]\mathbf{m/s}$ \\
        \midrule
        Left bumper  & Move     & Trap start command      & -\\
        \midrule
        Right bumper & Dribble  & Kick start command      & -\\
                     & (Pass)   & ground pass  & -\\
        \midrule
        Left stick   & Move     & Target move velocity    & $\left[0, 7\right]\mathbf{m/s}$ \\
                     & Dribble  & Target dribble velocity & $\left[0, 7\right]\mathbf{m/s}$ \\
        \midrule
        Right stick  & Move     & Target facing direction & -\\
                     & Kick     & Target kick direction   & $\left[0, 45\right]\mathbf{^\circ}$ \\
                     & (Pass)   & Pass target & - \\
        \midrule
        B button     & Kick     & Kick end command        & -\\
                     & Trap     & Trap end command        & -\\
        \midrule
        Directional pad& Move& Player switch&-\\ 
        \bottomrule
    \end{tabular}
    \label{tbl:gamepad}
\end{table}

\begin{figure}[H]
  \centering
  \includegraphics[trim=0 20 10 52, clip, width=0.7\linewidth]{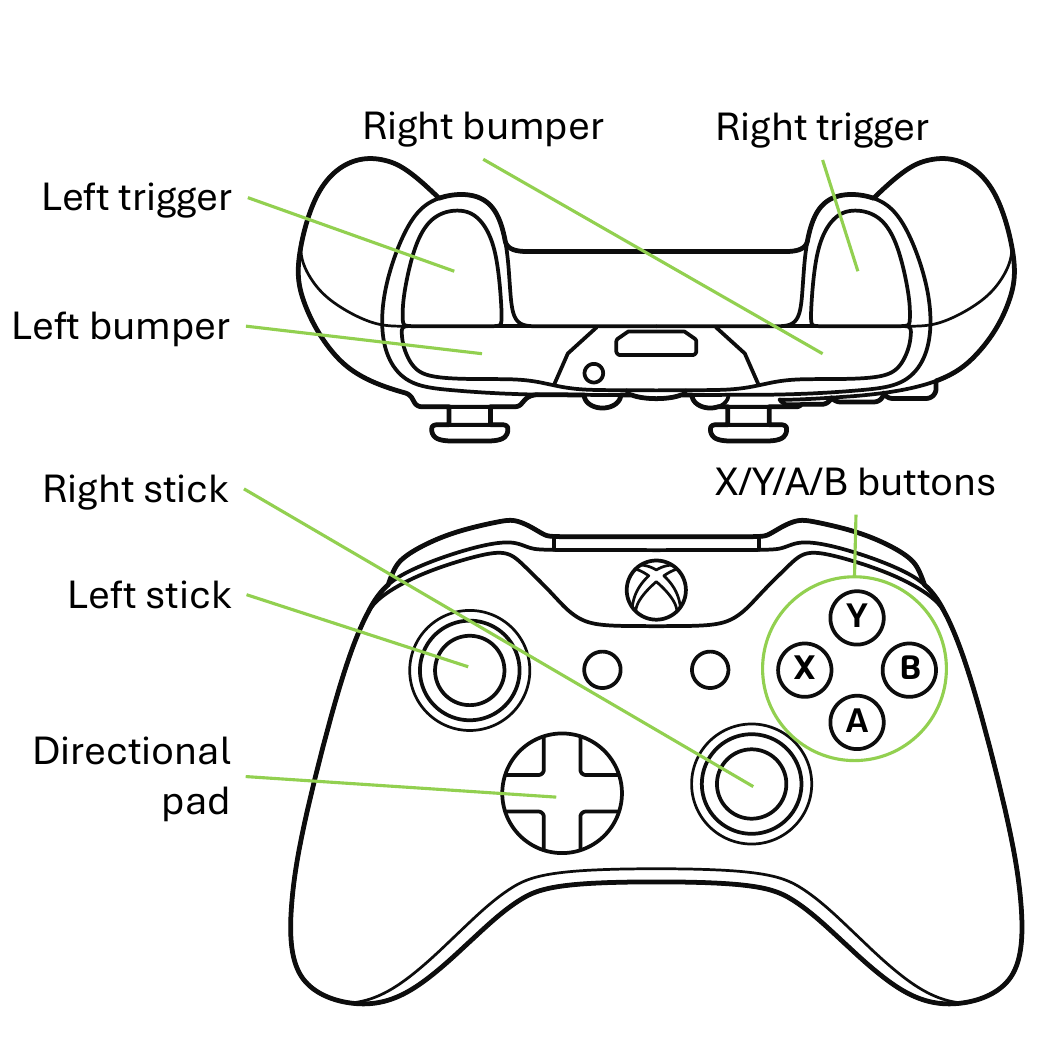}
  \caption{Gamepad inputs.}
  \label{fig:gamepad}
\end{figure}

The functions of each gamepad input in our interactive demos (Section~\ref{sec:interactive-demo}) are summarized in Table~\ref{tbl:gamepad} and Figure~\ref{fig:gamepad}. Pass-related inputs are enabled only in demos with at least one team player other than the controlled player. The Pass state is maintained when the Left Trigger input value exceeds 0.01.
In the Pass state, the default pass type is a Lob pass, but pressing the Right Bumper switches it to a Ground pass. The pass target is selected as the team player closest to the horizontal direction indicated by the Right Stick, relative to the controlled player's orientation.

\end{document}